\documentclass[aps,twocolumn,superscriptaddress,showpacs,floatfix,longbibliography]{revtex4-1}

\makeatletter
\def\@bibdataout@aps{%
\immediate\write\@bibdataout{%
@CONTROL{%
apsrev41Control%
\longbibliography@sw{%
    ,author="08",editor="1",pages="1",title="0",year="1"%
    }{%
    ,author="08",editor="1",pages="1",title="",year="1"%
    }%
  }%
}%
\if@filesw \immediate \write \@auxout {\string \citation {apsrev41Control}}\fi
}
\makeatother
\usepackage{graphicx}
\usepackage{dcolumn}
\usepackage{bm}
\usepackage{multirow}
\usepackage{booktabs}
\usepackage{array}

\usepackage{amssymb}
\usepackage{amsmath}
\usepackage{graphicx}
\usepackage{xcolor}

\usepackage[colorlinks,citecolor=blue,linkcolor=red,urlcolor=blue]{hyperref}
\usepackage{enumerate}

\begin{document}
\title{Relaxation Critical Dynamics with Emergent Symmetry}
\author{Yu-Rong Shu}
\affiliation{School of Physics and Materials Science, Guangzhou University, Guangzhou 510006, China}

\author{Ting Liao}
\affiliation{School of Physics and Materials Science, Guangzhou University, Guangzhou 510006, China}

\author{Shuai Yin}
\email{yinsh6@mail.sysu.edu.cn}
\affiliation{Guangdong Provincial Key Laboratory of Magnetoelectric Physics and Devices, Sun Yat-sen University, Guangzhou 510275, China}
\affiliation{School of Physics, Sun Yat-Sen University, Guangzhou 510275, China}

\date{\today}

\begin{abstract}
Universal critical properties can manifest themselves not only in \textit{spatial} but also in \textit{temporal} directions. It has been found that critical point with emergent symmetry exhibits intriguing spatial critical properties characterized by two divergent length scales, attracting long-term investigations. However, how the \textit{temporal} critical properties are affected by emergent symmetry is largely unknown. Here we study the nonequilibrium critical dynamics in the three-dimensional ($3$D) clock model, whose critical point has emergent $U(1)$ symmetry. We find that in contrast to the magnetization $M$, whose relaxation process is described by the usual dynamic exponent $z$ of the $3$D XY universality class, the angular order parameter $\phi_q$ shows a remarkable two-stage evolution characterized by different dynamic critical exponents. While in the short-time stage the relaxation dynamics is governed by $z$, in the long-time stage the dynamics is controlled by a new dynamic exponent $z'$. Further scaling analyses confirm that $z'$ is an indispensable dynamic critical exponent. Our results may be detected in the hexagonal RMnO$_3$ (R$=$rare earth) materials experimentally.
\end{abstract}

\maketitle

{\bf Introduction}---Symmetry is at the heart of identifying, characterizing and classifying phases and phase transitions, resulting in the profound notion of the universality class, which is one of organizing principles in modern condensed matter physics~\cite{Landaubook}. At a critical point, long wave-length fluctuations can gloss over the detailed information of the Hamiltonian, endowing the critical system with higher emergent symmetry than that of the original Hamiltonian. This emergent symmetry engenders plenty of intriguing critical phenomena and attracts enormous investigations in various systems ranging from classical to quantum phase transitions. For instance, in classical cases, the three dimensional ($3$D) $q$-state clock model exhibits $U(1)$ symmetry at the critical point for $q\geq 4$~\cite{Oshikawa2000prb,Lou2007prl,Okubo2015prb,Leonard2015prl,Pujari2015prb,Ding2016prb,Hasenbusch2011prb,Shao2020prl,Patil2021prb}. In quantum Dirac systems, the emergent $U(1)$ symmetry arises at the fermion-induced quantum critical point hosted in the Kekule VBS phase transition whose microscopic Hamiltonian has $Z_3$ discrete symmetry~\cite{Li2017nc,Jian2017prb,Jian2018prb,Torres2018prb,Classen2017prb}.
In quantum magnets, the most prominent example is the emergent $SO(5)$ symmetry appearing in deconfined quantum critical point~\cite{Senthil2004sci,Senthil2004prb,Nahum2015prl,Wang2017prx,Takahashi2020prr,Ma2019prl}.

A common feature for these critical points with emergent symmetry is that there exists a dangerously irrelevant scaling variable (DISV), which is irrelevant at the critical point,  but relevant in the ordered phases~\cite{Amit1982annph}. The DISV can induce a striking critical property in the \textit{spatial} direction: an extra divergent length scale $\xi'$ develops, in addition to the conventional correlation length $\xi$~\cite{Oshikawa2000prb}. Moreover, in sharp contrast to usual criticality dominated by single $\xi$, fertile exotic scaling behaviors spring up as a result of the interplay between $\xi$ and $\xi'$~\cite{Oshikawa2000prb,Lou2007prl,Okubo2015prb,Leonard2015prl,Pujari2015prb,Ding2016prb,Hasenbusch2011prb,Shao2020prl,Patil2021prb,Li2017nc,Jian2017prb,Jian2018prb,Senthil2004sci,Senthil2004prb,Nahum2015prl,Wang2017prx,Takahashi2020prr,Ma2019prl,Torres2018prb,Classen2017prb,Nelson1976prb,Amit1982annph,Miyashita1997jpsj,Shao2016Sci}.

However, how the emergent symmetry with DISV casts its light over the critical properties in the \textit{temporal} direction is still in the dark. This question is motivated by the fact that besides the static critical properties, symmetry and conserved quantities also play pivotal roles in characterizing nonequilibrium critical dynamics. In classical systems, different dynamic universality classes are classified according to the symmetric property of the order parameter and the coupling between the order parameter and the conserved secondary densities~\cite{Hohenberg1977rmp,Folk2006jpa,Tauber2014book}. Dictated by dynamic universality classes, nonequilibrium critical dynamics shows colorful universal time-dependent scaling behaviors and currently becomes one of the most vibrant issues in statistical physics~\cite{Hohenberg1977rmp,Folk2006jpa,Tauber2014book}.

Another motivation for this question is inspired by recent experimental research on nonequilibrium critical dynamics of $Z_q$-clock model fervently pursued in hexagonal RMnO$_3$ (R$=$rare earth) materials~\cite{Chae2012prl,Skjaervo2019prx,Griffin2012prx,Lin2014natphy,Meier2017prx,Meier2020prb,Zhang2021prb,OWSandvik2023nl,Kang2023jap,Baghizadeh2019jpcc,Juraschek2020prl}.
These experiments showed that the scaling of topological defects under external driving belongs to Model A with $3$D XY universality class, establishing the ``irrelevant'' aspect of the DISV in critical dynamics~\cite{Chae2012prl,Skjaervo2019prx,Griffin2012prx,Lin2014natphy,Meier2017prx}. However, an essential puzzle arises: how does the ``dangerous'' side of the DISV manifest itself in the nonequilibrium dynamics?

To seek for answers, here we study the dissipative nonequilibrium relaxation dynamics in $3$D $Z_q$-clock model starting from an ordered initial state. For the $3$D XY model, the magnitude and angular fluctuations share the same dynamic exponent $z=2.0246(10)$~\cite{Adzhemyan2022pa}.
For the $Z_q$-clock model, we show that the whole relaxation process of the magnitude of magnetization $M$ is described by the usual dynamic exponent $z$ governed by the dynamic universality class of the $3$D XY model, showing the irrelevant aspect of the DISV~\cite{Oshikawa2000prb,Lou2007prl,Okubo2015prb,Leonard2015prl,Pujari2015prb,Ding2016prb,Hasenbusch2011prb,Shao2020prl,Patil2021prb}.
In contrast, the dynamics of the angular order parameter $\phi_q$ is branded by the ``dangerous'' aspect of the DISV. A remarkable two-stage evolution is discovered for $\phi_q$. In the short-time stage, $\phi_q$ is a dimensionless variable and its dynamics is characterized by the $z$, whereas in the long-time stage $\phi_q$ gets a scaling dimension from the DISV and its dynamics is characterized by a new dynamic exponent $z'$ with $z'>z$. Moreover, we find that $z'$ increases with $q$. Further scaling analyses confirm that $z'$ is an indispensable critical exponent. Our present work not only sheds new light on the \textit{temporal} critical properties with emergent symmetry, but also brings new vitality into experimental investigations in the hexagonal RMnO$_3$ materials~\cite{Chae2012prl,Skjaervo2019prx,Griffin2012prx,Lin2014natphy,Meier2017prx,Meier2020prb,Zhang2021prb,OWSandvik2023nl,Kang2023jap,Baghizadeh2019jpcc,Juraschek2020prl}. Furthermore, our results may exert influence on nonequilibrium dynamics of supersolid quantum phase transitions in materials such as Na$_2$BaCo(PO$_4$)$_2$~\cite{Xiang2024nat,Chi2024,Gao2024}.

\begin{figure}[tbp]
\centering
  \includegraphics[width=\linewidth,clip]{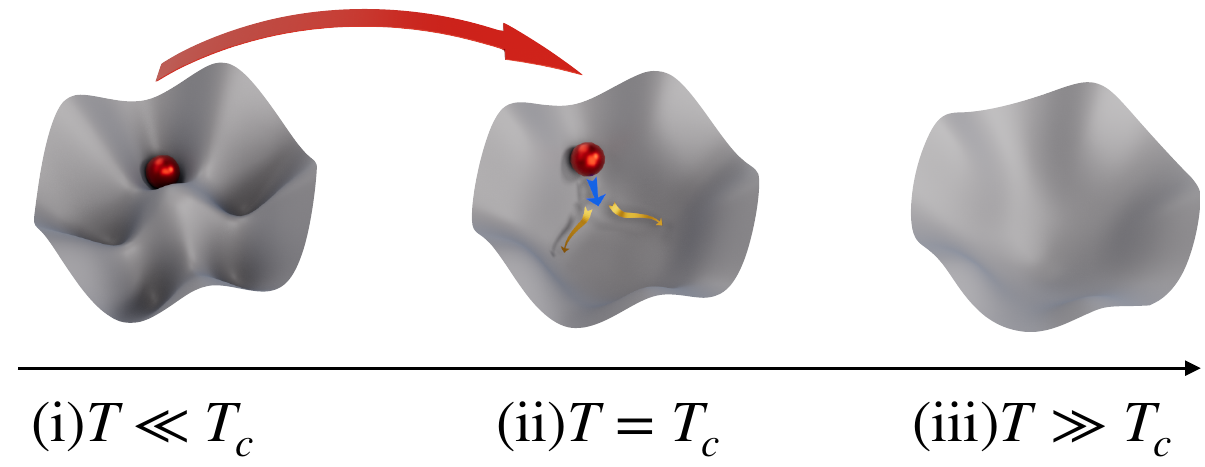}
  \vskip-3mm
  \caption{Sketch of free energy landscape for different temperatures: (i) ordered state at $T\ll T_c$ with the puddles representing the five possible directions of $\theta$; (ii) critical state at $T_c$ with shallow minimum region in the free energy; (iii) disordered state at $T\gg T_c$ with a deep minimum in the free energy.
    The red arrow indicates a quench from an ordered state (i) to the critical temperature (ii).
    The red sphere stands for an ordered state with all spins pointing in the same chosen direction in (i) and the relaxation after the quench in (ii). The blue arrow in (ii) indicates an initial decrease of the magnitude of the order parameter and the yellow arrows suggest the angular dispersion.
  }
  \label{fig:quench}
\end{figure}

{\bf Model and numerical method}---The Hamiltonian of the ``hard'' $Z_q$-clock model in cubic lattice is~\cite{Oshikawa2000prb,Lou2007prl,Okubo2015prb,Leonard2015prl,Pujari2015prb,Ding2016prb,Hasenbusch2011prb,Shao2020prl,Patil2021prb}
\begin{equation}
\label{eq:hamiltonian}
H=-\sum_{\langle ij\rangle}\cos(\theta_i-\theta_j),
\end{equation}
in which $\theta=2n\pi/q$ with $n\in \{0,\cdots,q-1\}$. The summation is taken among nearest pairs denoted by $\langle\rangle$. In $3$D this model has a direct phase transition between ordered and disordered phase at $g\equiv T-T_c=0$, as illustrated in Fig.~\ref{fig:quench}. The discreteness of $\theta$ implies a Z$_q$ field which serves as a hidden DISV and an emergent $U(1)$ symmetry appears at $T_c$. An explicit DISV can be revealed in a ``soft'' version of (\ref{eq:hamiltonian}) where the angles vary continuously in $[0,2\pi)$, with an additional discrete symmetric term $-h\sum_i \cos(q \theta_i)$ serving as the DISV.
When $q\geq 5$ for the hard model and $q\geq 4$ for the soft model, the phase transition belongs to the $3$D XY universality class~\cite{Oshikawa2000prb,Lou2007prl,Okubo2015prb,Leonard2015prl,Pujari2015prb,Ding2016prb,Hasenbusch2011prb,Shao2020prl,Patil2021prb}.
However, when $g<0$ the discreteness becomes relevant and the ordered phase breaks $Z_q$ symmetry.
In addition to correlation length $\xi$ with the $3$D XY exponent $\nu=0.6717(1)$~\cite{Campostrini2006prb,Campostrini2001prb,Chester2020jhep},
the critical point of model~(\ref{eq:hamiltonian}) has another typical length scale $\xi'$ related to the effective mass induced by DISV~\cite{Ueno1991prb,Chubukov1994prb,Oshikawa2000prb,Leonard2015prl,Banerjee2018prl}, and its associated critical exponent is $\nu'$ which is larger than $\nu$ and increases with $q$~\cite{Lou2007prl,Okubo2015prb,Leonard2015prl,Pujari2015prb,Ding2016prb,Hasenbusch2011prb,Shao2020prl,Patil2021prb,Ueno1991prb,Chubukov1994prb,Banerjee2018prl}.

Here we study the relaxation dynamics starting from an ordered initial state to the vicinity of the critical point at $t=0$, as shown in Fig.~\ref{fig:quench}. The Monte Carlo method with standard Metropolis dynamics is used to simulate the nonequilibrium evolution~\cite{binderbook}. The time unit is defined as a full Monte Carlo sweep through the lattice. It has been shown that this kind of Monte Carlo dynamics belongs to the Model A universality class~\cite{Tauber2014book}, and is straightforward to implement in experiments~\cite{Chae2012prl,Skjaervo2019prx,Griffin2012prx,Lin2014natphy,Meier2017prx}.

{\bf General scaling theory}--- For the dynamics of a general physical quantity $P$ with dimension $\kappa$, we propose a reformed dynamic scaling form
\begin{equation}
\label{eq:generalscaling}
P(t,g,L)=L^{-\kappa}f_P(tL^{-z},tL^{-z'};gL^{1/\nu},gL^{1/\nu'}),
\end{equation}
in which a new dynamic exponent $z'$ is discovered, in addition to the usual dynamic exponent $z$, to take into account the effects induced by DISV. In the absence of DISV, Eq~(\ref{eq:generalscaling}) reduces to the usual scaling form without $tL^{-z'}$ and $gL^{1/\nu'}$, which has been verified by both theoretical~\cite{Fedorenko2006epl} and numerical methods~\cite{Zhengb1998ijmpb,Luo1998prl,Stauffer1992pa,Ito1993pa,Albano2011rpp}. In addition, when $t\rightarrow\infty$, the equilibrium scaling form with two correlation-length exponents is recovered~\cite{Okubo2015prb,Leonard2015prl,Shao2020prl,Patil2021prb}.

We explore the evolution of the square of order parameter $M^2=\langle M_x^2\rangle +\langle M_y^2 \rangle$, in which $\langle\rangle$ stands for statistical average. The components are defined as $M_x\equiv \sum_{i} \cos(\theta_i)/L^d$ and $M_y\equiv \sum_{i} \sin(\theta_i)/L^d$.
We also particularly focus on the dynamics of the angular order parameter $\phi_q$ defined as $\phi_q=\langle\cos(q \Theta)\rangle$ with $\Theta\equiv \arccos(M_x/M)$. Note that $\phi_q$ characterizes the discrete symmetry breaking and becomes nonzero in response to the $Z_q$ field. In equilibrium, the scaling of $\phi_q$ directly reflects the impact of $\xi'$~\cite{Lou2007prl,Okubo2015prb,Pujari2015prb,Shao2020prl,Patil2021prb}.
In the following, without loss of generality, we will study both $q=5$ and $q=6$ hard clock model~(\ref{eq:hamiltonian}).

\begin{figure*}[!htbp]
\centering
  \includegraphics[width=\linewidth,clip]{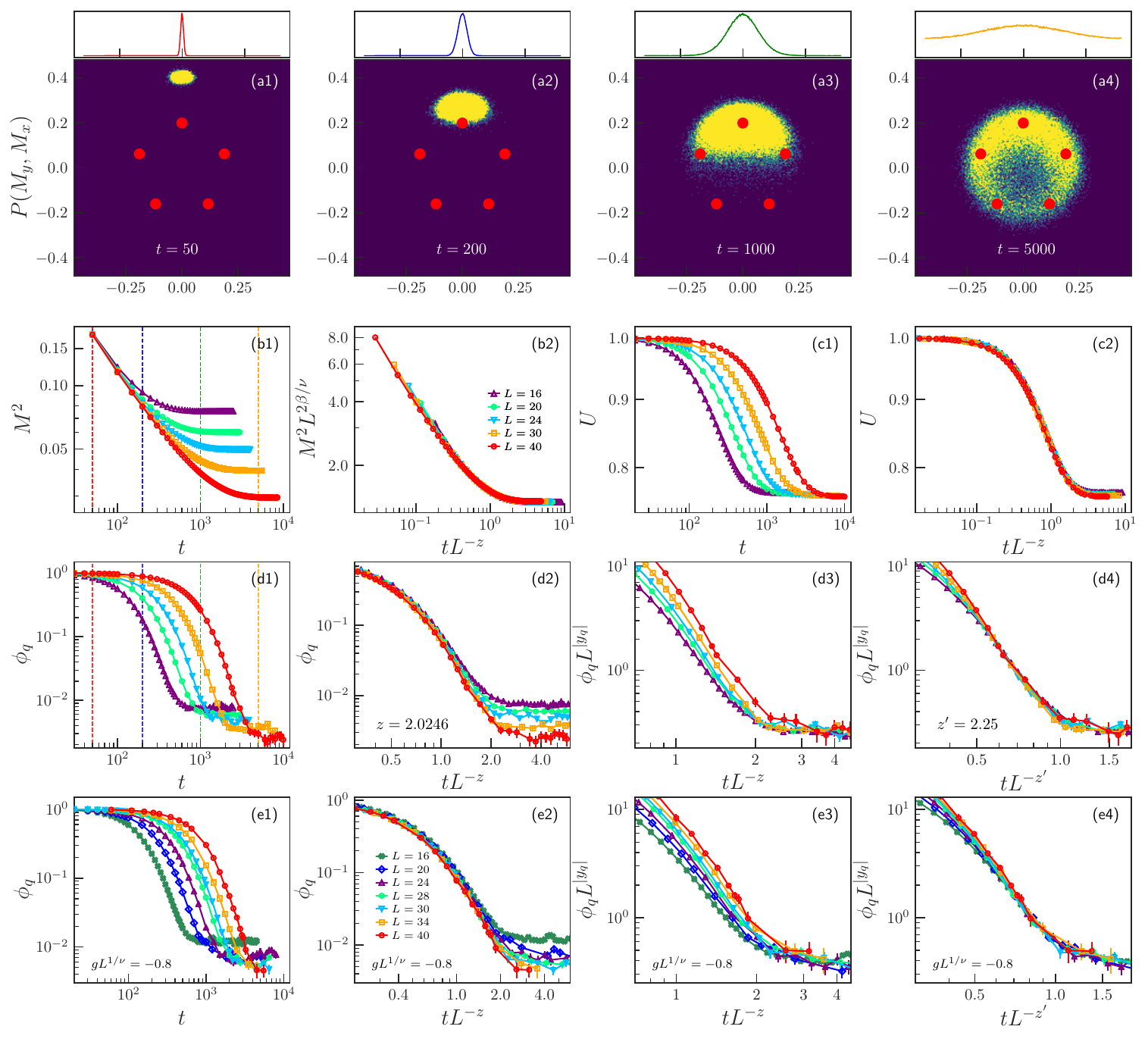}
  \vskip-3mm
  \caption{Relaxation dynamics of the clock model with $q=5$.
    (a1)-(a4): Distribution of $M_x$ and $M_y$ for four typical time values characterizing different evolution stages with $L=30$. The five red dots serve as a visual guide, showing the equilibrium value of $M$ and the angular discreteness. Angular distributions $P(\theta)$ with $\theta=[0,2\pi)$ are shown above each histogram.
    (b1): Time dependence of $M^2$ for $L=16$ to $40$. The dashed vertical lines which correspond to the time value in (a1)-(a4), respectively, are shown to indicate different evolution stages.
    (b2): After rescaling $M^2$ and $t$ as $M^2L^{2\beta/\nu}$ and $tL^{-z}$, all curves collapse well.
    (c1)-(c2): Evolution of Binder cumulant $U$ before and after rescaling $t$ as $tL^{-z}$.
    (d1): Time dependence of $\phi_q$. The dashed vertical lines are the same as (b1).
    (d2): At early stage of evolution, $\phi_q$ is dimensionless and $t$ is rescaled by $tL^{-z}$.
    (d3): At long times, no good collapse is seen when $\phi_q$ is rescaled by $\phi_q L^{|y_q|}$ and $t$ by $tL^{-z}$.
    (d4): Similar to (d3) but $t$ is rescaled by $tL^{-z'}$ with $z'=2.25$. The good collapse indicates the dynamics is governed by $z'$ instead of $z$.
    (e1)-(e4): Scaling analyses similar to (d1)-(d4) but for off-critical-point case with fixed $gL^{1/\nu}=-0.8$.
  }
  \label{fig:figure2}
\end{figure*}
\begin{figure*}[!htbp]
\centering
  \includegraphics[width=\linewidth,clip]{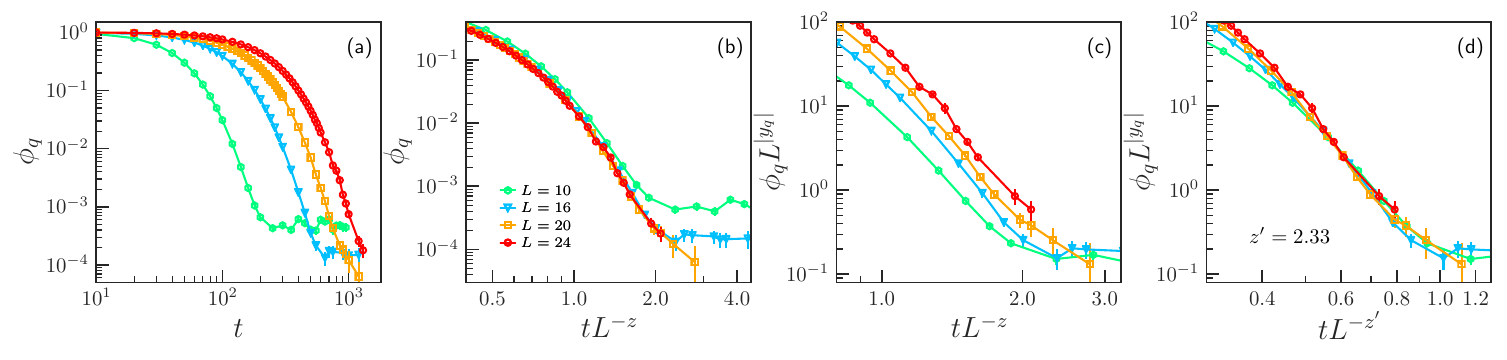}
  \vskip-3mm
  \caption{Relaxation dynamics of the clock model with $q=6$.
    (a): Time dependence of $\phi_q$.
    (b): At short-time stage of evolution, $\phi_q$ is dimensionless and $t$ is rescaled by $tL^{-z}$.
    (c): At long times, no good collapse is seen when $\phi_q$ is rescaled by $\phi_q L^{|y_q|}$ and $t$ by $tL^{-z}$.
    (d): Rescaling $t$ by $tL^{-z'}$ with $z'=2.33$ makes the rescaled curves collapse well, indicating the long-time dynamics is governed by $z'$ instead of $z$.
  }
  \label{fig:figure3}
\end{figure*}

{\bf Case of} $\bm{q=5}$---We first study the relaxation dynamics for $q=5$ at its critical point $T_c=2.20502(1)$~\cite{Shao2020prl} quenching from an ordered state with $M_x=1$ as indicated in Fig.~\ref{fig:quench}. Figures~\ref{fig:figure2}(a1)-(a4) show the evolution of the distribution of $M_x$ and $M_y$. In early stage, the magnitude of $M$ (represented by the radius) decays as time increases, with very limited angular dispersion.
In later stage, the magnitude is almost invariant, whereas the transverse fluctuations dominate and the angular distribution spreads until the system reaches its equilibrium state, in which a circular distribution forms, reflecting the emergent $U(1)$ symmetry. Note that Fig.~\ref{fig:figure2}(a4) only shows an approximate $U(1)$ symmetry since the emergent symmetry only appears in the thermodynamic limit.

We first show the ``irrelevant'' aspect of the DISV in relaxation dynamics. The evolution of $M^2$ is shown in Fig.~\ref{fig:figure2}(b1) for various system sizes. By rescaling $M^2$ and $t$ as $M^2 L^{2\beta/\nu z}$ and $t L^{-z}$, respectively, in which $\beta=0.3486(1)$, $\nu=0.6717(1)$~\cite{Campostrini2006prb,Campostrini2001prb,Chester2020jhep}, and $z=2.0246(10)$~\cite{Adzhemyan2022pa} are all $3$D XY critical exponents, we find that the rescaled curves collapse well, as shown in Fig.~\ref{fig:figure2}(b2), suggesting that the dynamics of $M^2$ obeys $M^2(t,L)=L^{-2\beta/\nu}f_M(tL^{-z})$. Similarly, the dimensionless Binder cumulant $U\equiv 2-\langle M\rangle^4/\langle M^2\rangle^2$ satisfies $U=f_U(tL^{-z})$ in the whole relaxation process, as confirmed in Figs.~\ref{fig:figure2}(c1)-(c2). These results demonstrate that emergent $U(1)$ symmetry can manifest itself in the nonequilibrium process, consistent with the experiments~\cite{Chae2012prl,Skjaervo2019prx,Griffin2012prx,Lin2014natphy,Meier2017prx}.

However, that is not the full picture. More strikingly, we here reveal the ``dangerous'' side of DISV in nonequilibrium relaxation dynamics by investigating the evolution of the angular order parameter $\phi_q$. We find that the relaxation process of $\phi_q$ exhibits a remarkable two-stage evolution.

In the short-time stage, $\phi_q$ is close to $1$. Thus it should be dimensionless as learned from the equilibrium scaling~\cite{Shao2020prl,Patil2021prb}. According to Eq.~(\ref{eq:generalscaling}), the dynamic scaling form of $\phi_q$ reads $\phi_q(t,L)=f_{\phi}^s(tL^{-z},tL^{-z'})$.
By rescaling $t$ as $tL^{-z}$, we find good collapse at short times, suggesting that $tL^{-z}$ dominates the scaling function $f_{\phi}^s$, as shown in Fig.~\ref{fig:figure2}(d2).
Therefore, in the short-time stage, $\phi_q$ satisfies
\begin{equation}
\label{eq:phiqlarge}
\phi_q(t,L)=f_{\phi}^s(tL^{-z}).
\end{equation}

In contrast, in the long-time stage, the relaxation of $\phi_q$ apparently deviates from Eq.~(\ref{eq:phiqlarge}) as shown in Fig.~\ref{fig:figure2}(d2).
A distinct scaling form of $\phi_q$ for long times is required. It has been found that in equilibrium, for finite-size systems discrete $Z_q$ symmetry makes $\phi_q$ obey $\phi_q\propto L^{-|y_q|}$ with $y_q$ the scaling dimension of DISV~\cite{Shao2020prl}.
Considering Eq.~(\ref{eq:generalscaling}), one expects that the dynamic scaling form should be $\phi_q\propto L^{-|y_q|}f_{\phi}^l(tL^{-z},tL^{-z'})$.
To make out the dominant variable in $f_{\phi}^l$, $\phi_q$ is rescaled by $\phi_qL^{|y_q|}$ with $y_q=-1.27(1)$ for $q=5$~\cite{Shao2020prl}, and for $t$, both cases of $tL^{-z}$, $tL^{-z'}$ are shown for comparison in Figs.~\ref{fig:figure2}(d3) and (d4), respectively. Strikingly, we find that $tL^{-z'}$ with $z'= 2.25$, rather than $tL^{-z}$, can successfully collapse all the rescaled curves at long times, demonstrating the long-time scaling form of $\phi_q$ should be
\begin{equation}
\label{eq:phiqsmall}
\phi_q(t,L)=L^{-|y_q|}f_{\phi}^l(tL^{-z'}).
\end{equation}
Different from the critical dynamics observed in the $3$D XY universality class, where only a single dynamic exponent $z$ is required~\cite{Tauber2014book}, here a new dynamic exponent $z'$ emerges, indicating the relevance of DISV in the relaxation process.

Moreover, we explore off-critical-point effects in the vicinity of the critical point. Scaling collapses in Figs.~\ref{fig:figure2}(e1)-(e4) show that for $gL^{1/\nu}=-0.8$, $\phi_q$ obeys $\phi_q(t,L)=f_{\phi}^s(gL^{1/\nu},tL^{-z})$ in the short-time stage, whereas in the long-time stage, $\phi_q$ obeys $\phi_q(t,L)=L^{-|y_q|}f_{\phi}^l(gL^{1/\nu},tL^{-z'})$ in which $tL^{-z'}$ with $z'=2.25$, rather than $tL^{-z}$, governs. These results not only generalize the scaling form to off-critical-point region (for more general cases, see Appendices B, C) but also confirm the value of $z'$ for $q=5$.

{\bf Case of} $\bm{q=6}$---To delve deeper into the universal temporal scaling properties involving emergent symmetry, we now turn to study relaxation dynamics for model~(\ref{eq:hamiltonian}) with $q=6$ at its critical point $T_c=2.20201(1)$~\cite{Shao2020prl}. In analogy to the $q=5$ case, $M^2$ and $U$ are governed by the usual $z$ (not shown), while $\phi_q$ exhibits the two-stage character. Figure~\ref{fig:figure3} confirms that in the short-time stage, $\phi_q$ is described by Eq.~(\ref{eq:phiqlarge}) in which $z$ also dominates; whereas in the long-time stage, $\phi_q$ is described by Eq.~(\ref{eq:phiqsmall}) with $y_q=-2.55(6)$~\cite{Shao2020prl} and the new dynamic exponent $z'=2.33$ dominates, verifying the universality of Eq.~(\ref{eq:phiqsmall}) and confirming that $z'$ is an indispensable critical exponent that plays essential roles in the nonequilibrium dynamics near the critical point. Moreover, we find that $z'$ depends on $q$ and its value for $q=6$ is larger than that for $q=5$.

\begin{table}[b]
  \caption{Two-stage dynamic scaling behavior of $\phi_q$ necessitates $z'$ as an indispensable critical exponent.}
  \label{tablecom}
  \begin{ruledtabular}
    \begin{tabular}{cccc}
  $\phi_q$ & Dimension & Equilibrium & Nonequilibrium \\
  \hline
  large & 0 & $\phi_q=f(gL^{1/\nu'})$ & $\phi_q= f(tL^{-z})$ \\
      small & $y_q$ & $\phi_q= L^{-|y_q|}f(gL^{1/\nu})$ & $\phi_q= L^{-|y_q|} f(tL^{-z'})$ \\
     \end{tabular}
  \end{ruledtabular}
\end{table}

{\bf Discussion}---Here we point out that the new dynamic exponent $z'$ is an indispensable dynamic critical exponent to analytically connect two scaling limits where $\phi_q$ is dimensionless when it is close to $1$ and $\phi_q\propto L^{-|y_q|}$ near equilibrium, as illustrated in Table~\ref{tablecom}. The reason is that if there were only one single dynamic exponent (assuming that it is $z$ without loss of generality) in both Eqs.~(\ref{eq:phiqlarge}) and (\ref{eq:phiqsmall}), Eq.~(\ref{eq:phiqlarge}) cannot convert to Eq.~(\ref{eq:phiqsmall}) analytically since $|y_q|$ in front of the scaling function $f_{\phi}^l$ cannot come out of thin air. As a consequence, an additional $z'$, which combines $z$ and $|y_q|$, must be introduced. This is quite similar to the equilibrium case in which $\nu'$ is required to build a bridge across the two cases of large and small $\phi_q$~\cite{Shao2020prl}, as compared in Table~\ref{tablecom} (See also Appendix A for detailed discussion).

Physically, from Figs.~\ref{fig:figure2}(a), (b1) and (d1), one finds that in the short-time stage longitudinal fluctuations govern the relaxation dynamics whereas the transverse fluctuations are localized in the vicinity of the initial angle. In this stage, the system is unable to distinguish between the $U(1)$ and $Z_q$ symmetry, such that the scaling behaviors are characterized by the usual dynamic exponent $z$.
In contrast, in the long-time stage, the transverse modes spread across the barriers between the minima of the free energy, which are absent in the 3D XY model, so that the system can perceive the difference and leads to the emergence of the new dynamic exponent $z'$.
Note that in this stage, the magnitude of $M$ remains almost unchanged, making quantities that mainly depends on $M$ be fully characterized by $z$. However, $z'$ emerges in the long-time dynamics of $\phi_q$, as shown in Figs.~\ref{fig:figure2}(d), (e) and \ref{fig:figure3}(c), since $\phi_q$ describes the angular fluctuations in the presence of DISV.

{\bf Summary and outlook}---To summarize, we have unveiled the \textit{temporal} critical properties with emergent symmetry by studying the nonequilibrium dynamics in $3$D $Z_q$-clock model. Using Monte Carlo simulation for cases of $q=5$ and $6$, we have found both the ``irrelevant' and ``dangerous'' aspects of the DISV in the relaxation critical dynamics. In particular, a new dynamic exponent $z'$ has been discovered to characterize the long-time dynamics of the angular order parameter.

Our present work undoubtedly provides new ingredients in recent experiments on the nonequilibrium critical dynamics in the hexagonal RMnO$_3$ materials, wherein the system has $Z_6$ symmetry and its critical point has $U(1)$ symmetry~\cite{Chae2012prl,Skjaervo2019prx,Griffin2012prx,Lin2014natphy,Meier2017prx,Meier2020prb,Zhang2021prb,OWSandvik2023nl,Kang2023jap,Baghizadeh2019jpcc,Juraschek2020prl}. Moreover, the increase of $z'$ with $q$ may be examined experimentally in Nd$_2$SrFe$_2$O$_7$~\cite{Huang2021prr} and Tb$_2$SrAl$_2$O$_7$~\cite{Xu2022cm},
in which phase transition with discrete $Z_8$-symmetry breaking happens. Our results also shed light on the dynamics of quantum criticality with emergent symmetry. For instance, the spin supersolid quantum phase transition with discrete $Z_6$ symmetry breaking in Na$_2$BaCo(PO$_4$)$_2$ has recently been explored~\cite{Xiang2024nat}. In addition, the fast-developing quantum annealer also provides promising platforms to study the nonequilibrium critical dynamics with emergent symmetry~\cite{Ali2024}. Our findings can be generalized in these systems. Moreover, our results can also inspire investigations on the nonequilibrium dynamics in fermion-induced quantum criticality and deconfined quantum criticality which go beyond the conventional Landau paradigm~\cite{Shu2022prl,Shu2022prb,Shu2023kz}.

{\bf Acknowledgements}---We thank A. W. Sandvik for helpful discussions. S.Y. is supported by the National Natural Science Foundation of China (Grants No. 12222515 and No. 12075324) and the Science and Technology Projects in Guangdong Province (Grants No. 2021QN02X561). Y.R.S. and T.L. are supported by the National Natural Science Foundation of China (Grant No. 12104109), the Science and Technology Projects in Guangzhou (Grants No. 202201020222 and 2024A04J2092).

\appendix
\setcounter{equation}{0}
\renewcommand\theequation{A\arabic{equation}}
\setcounter{figure}{0}
\renewcommand\thefigure{A\arabic{figure}}

\section{General scaling form}
\label{Sec1}
In the main text, we propose the general dynamic scaling form for a quantity $Q$ with its critical dimension $\kappa$. Here we give a detail scaling analysis. Before introducing the scaling form with dangerously irrelevant scaling variable in Sec.~\ref{sec12}, we revisit the scaling theory with two relevant parameters in usual critical point for comparison in Sec.~\ref{sec11}.

\subsection{\label{sec11}Two-parameter scaling form in usual critical point}

We first give a brief review of the scaling form including two relevant scaling variables near a usual critical point $T_c$. Besides the distance to the critical point, $g\equiv T-T_c$, we assume that in the Hamiltonian, there exists a symmetry breaking term $-h\sum_i S_i$, with $S_i$ being the field of local order parameter. Since {\textbf{both $g$ and $h$ represent the relevant directions}}, the scaling form of $Q$ for a typical momentum $k$ reads
\begin{equation}
  Q(g,h,k)=k^{\kappa}f_{Q1}(kg^{-\nu},kh^{-\nu/\beta\delta}),
  \label{eq:usufull}
\end{equation}
in which $\delta$ is defined according to the scaling relation between the order parameter $M$ and $h$, i.e., $M\propto h^{1/\delta}$ at $g=0$. Note that here both $g$ and $h$ are {\textbf{independent}} relevant scaling variables. Phase transition occurs when $g=0$ and $h=0$.

For finite size system with linear size $L$, $L^{-1}$ provides a typical momentum. In this way, the usual finite-size scaling is restored. In particular, for the square order parameter $M$, Eq.~(\ref{eq:usufull}) gives
\begin{equation}
  M^2(g,h,L)=L^{-2\beta/\nu}f_{M1}(gL^{1/\nu},hL^{\beta\delta/\nu}).
  \label{eq:usuop}
\end{equation}

For the nonequilibrium relaxation dynamics, when the initial state corresponds to the fixed point of the scaling transformation, such as a completely ordered state, the scaling form of the order parameter reads~\cite{Tauber2014book,Landau2009book,Janssen1989,Li1995prl}

\begin{equation}
  M(g,h,L,t)=L^{-\beta/\nu}f_{M2}(gL^{1/\nu},hL^{\beta\delta/\nu},tL^{-z}),
  \label{eq:usuop1}
\end{equation}
in which $z$ is the dynamic exponent. This scaling form has been extensively studied and efficient numerical methods have been developed based on it to determine the critical point and critical exponents during the nonequilibrium relaxation process in various systems~\cite{Tauber2014book,Landau2009book,Janssen1989,Li1995prl}.

\subsection{\label{sec12}Scaling form in $Z_q$ clock model in the presence of dangerously irrelevant scaling variable}
\subsubsection{\label{eq}Equilibrium scaling forms}

Now, let us turn to our main discoveries in the $Z_q$ clock model with the $XY$ Hamiltonian accompanied by an additional {\textbf{dangerously irrelevant}} term $\lambda_q (\Phi^q+\Phi^{*q})$ ($q\geq 4$) (here $\Phi$ denotes the complex order parameter field). This term is the lowest order term breaking the symmetry from $U(1)$ to $Z_q$. The corresponding Euclidean action is just Eq.~(3) in Ref.~\cite{Oshikawa2000prb}. It was shown that $\lambda_q$ is irrelevant at the critical point~\cite{Amit1982annph,Oshikawa2000prb}. Accordingly, the critical point has an emergent $U(1)$ symmetry. Thus, it looks like that the critical phenomena for this model could be described by the critical theory of the $XY$ model. If so, the scaling form of a quantity $Q$ should be
\begin{equation}
  Q(g,\lambda_q,k)=k^{\kappa}f_{Q2}(gk^{-1/\nu},\lambda_qk^{|y_q|}),
  \label{eq:usufull1}
\end{equation}
in which $y_q$ the critical dimension of $\lambda_q$ and it is negative. Accordingly, $\lambda_q$ only plays minor roles like the scaling correction.

However, $\lambda_q$ is \textbf{not} a usual irrelevant scaling variable, but a {\textbf{dangerously irrelevant scaling variable}}. In other words, at $T=T_c$, $\lambda_q$ is irrelevant; when $T<T_c$, $\lambda_q$ becomes relevant immediately, driving the ordered phase spontaneously breaking the discrete $Z_q$ symmetry. It was demonstrated that the scaling behaviors in the ordered side (with $T$ still close to $T_c$) of this model cannot be fully captured by Eq.~(\ref{eq:usufull1}) in which the relevant side of $\lambda_q$ is completely ignored. Meanwhile, scaling form with two independent relevant variables, similar to Eq.~(\ref{eq:usufull}), also cannot be used to describe the scaling properties with $\lambda_q$ term, since the irrelevant side of $\lambda_q$ is not considered.

Accordingly, a proper scaling form should be able to reflect both relevant and irrelevant sides of DISV, with the following constraints to fulfill:

\begin{itemize}
\item {\color{blue}{Constraint (a): The limit of $\lambda_q\rightarrow 0$}} --- Without presence of $\lambda_q$, the scaling form should restore the usual scaling form of the $XY$ model.

\item {\color{blue}{Constraint (b): Relevant side of $\lambda_q$}} --- A relevant term including $\lambda_q$ must be included in the scaling form when $T<T_c$; But when $T=T_c$, in the limit of $L\rightarrow\infty$ and $k\rightarrow 0$, Eq.~(\ref{eq:usufull1}) should be recovered. This constraint indicates that this relevant term must {\textbf{not}} be the single $\lambda_q$ (similar to usual independent relevant variable like the symmetry breaking field $h$ as shown in Eq.~(\ref{eq:usufull})), but rather a composite variable which combining $g$ with $\lambda_q$.

\item {\color{blue}{Constraint (c): Irrelevant side of $\lambda_q$}} --- For any fixed finite $\lambda_q$, phase transition can still occurs and the scaling form should keep the same. This is different from the case of Eq.~(\ref{eq:usufull}), in which for finite $h$ phase transition cannot occur.

\end{itemize}

These constraints give the general scaling form
\begin{equation}
  Q(g,\lambda_q,k)=k^{\kappa}f_{Q3}[kg^{-\nu},k\mathcal{F}_q(g,\lambda_q),k\lambda_q^{1/|y_q|}],
\end{equation}
in which $\mathcal{F}_q(g,\lambda_q)$ should be a product function of $g$ and $\lambda_q$ to guarantee that $\lambda_q$ plays relevant roles only for $g\neq 0$. Analyses based on the functional renormalization group in Ref.~\cite{Leonard2015prl} showed that $\mathcal{F}_q(g,\lambda_q)\sim g^{-\nu(1+|y_q|/2)}\lambda_q^{-1/2}$ which comes from the renormalization flow of the transverse mass $m_T^2$. For example, for $q=6$, $m_T^2\propto \lambda_6 \Phi_{\textrm{min}}^4$~\cite{Leonard2015prl}.

Several remarks on this scaling form are outlined below:

\begin{enumerate}[(1)]

\item \label{remark1} The first variable $kg^{-\nu}$ in $f_{Q3}$ identifies the usual correlation length $\xi$ as $\xi^{-1}\sim k_\xi$ for $k_\xi g^{-\nu}=1$.
  
\item \label{remark2}  The second variable $kg^{-\nu(1+|y_q|/2)}\lambda_q^{1/2}$, which stems from the renormalization of the transverse mass $m_T^2$, identifies another length scale $\xi'$ as $\xi^{-1}\sim k_{\xi'}$ for $k_{\xi'} \sim g^{-\nu'}\lambda_q^{1/2}=1$ with $\nu'=\nu(1+|y_q|/2)$.
\item The third term $k\lambda_q^{1/|y_q|}$ now can be regarded as a scaling correction term, which can be neglected when the leading contributions are considered.
\item When $\lambda_q\rightarrow 0$, the usual scaling function of $XY$ model can be recovered, satisfying {\color{blue}{Constraint (a)}}.

\item The form of the second variable satisfies {\color{blue}{Constraint (b)}}.

\item Moreover, for a given specific system, the value of $\lambda_q$ is fixed. We assume $\lambda_q=C$. Thus {\color{blue}{Constraint (c)}} dictates that the scaling form for this specific system should be
\begin{equation}
  Q(g,k)=k^{\kappa}f_{Q4}(kg^{-\nu},Ckg^{-\nu'}).
  \label{eq:emefull4}
\end{equation}

\end{enumerate}

Comparing Eq.~(\ref{eq:emefull4}) with Eq.~(\ref{eq:usufull}), one finds that a fixed the symmetry breaking field $h$ can independently signify the departure from the critical point in the independent direction different from $g$, giving the two-parameter scaling form. In contrast, here, $\lambda_q$ cannot undertake an independent relevant direction. In this way, Eq.~(\ref{eq:emefull4}) retreats back to a one-parameter scaling with $g$ being the only relevant direction, but with two length scales, which are the usual correlation length scale and the other scale related to the transverse mass, respectively.

Next, we show how Eq.~(\ref{eq:emefull4}) can characterize the complex critical behaviors for different quantities. For the square of the order parameter $M^2$, it was shown that the finite-size scaling is the same as the usual one, i.e., Eq.~(\ref{eq:usuop}) with $h=0$~~\cite{Lou2007prl}. Another quantity, the angular order parameter $\phi_q$, defined as $\phi_q=\langle\cos(q \Theta)\rangle$ with $\Theta\equiv \arccos(M_x/M)$, has been widely used to reflect the role of discrete symmetry term, since $\phi_q=0$ in the absence of this DISV term. Recently, it was shown that the scaling form for $\phi_q$ is~\cite{Shao2020prl}
\begin{equation}
  \phi_q(g,L)=L^{-|y_q|}f_{\phi}(gL^{1/\nu},gL^{1/\nu'}).
  \label{eq:equiphi1}
\end{equation}
For $g=0$, it reduces to the scaling relation
\begin{equation}
  \phi_q(g,L)\propto L^{-|y_q|}.
  \label{eq:equiphi2_1}
\end{equation}
Our results in the very long-time equilibrium stage are consistent with this scaling relation (See Figs. 2, 3 for $q=5$ and $6$, respectively, in the main text). This scaling relation demonstrates that for finite-size systems the discrete symmetry can even manifest itself to the critical point. The reason is that the emergent symmetry is the property of the infrared fixed point which corresponds to the equilibrium critical state in thermodynamic limit. For finite $L$, residual information of the discrete symmetry can still exist at $T_c$. This phenomenon is similar to that the square of the order parameter $M^2$ for $g=0$ obeys $M^2\propto L^{-2\beta/\nu}$, which also reflects the residual information of the ordered phase for finite-size system.

Next we discuss the properties of $f_{\phi}(gL^{1/\nu},gL^{1/\nu'})$ in detail. In Ref.~\cite{Shao2020prl}, it was shown that for small $g$, the first variable of $f_{\phi}$ in Eq.~(\ref{eq:equiphi1}), i.e., $gL^{1/\nu}$, dominates and the scaling form becomes
\begin{equation}
  \phi_q(g,L)=L^{-|y_q|}f_{\phi1}(gL^{1/\nu}).
  \label{eq:equiphi2}
\end{equation}
As $g$ grows larger, the second variable starts to play roles, and Eq.~(\ref{eq:equiphi1}) demonstrates a crossover behavior
\begin{equation}
  \phi_q(g,L)=L^{-|y_q|}(gL^{1/\nu})^af_{\phi2}(gL^{1/\nu'}),
  \label{eq:equiphi3}
\end{equation}
in which $a=\nu(2+|y_q|)$ is a crossover exponent. For large $g$ (but still in the critical region) and $L$, $\phi_q\rightarrow 1$. This requires that $f_{\phi2}(gL^{1/\nu'})\sim (gL^{1/\nu'})^{-a}$, and the scaling form reduces to
\begin{equation}
  \phi_q(g,L)=f_{\phi3}(gL^{1/\nu'}).
  \label{eq:equiphi4}
\end{equation}

Here are several remarks on the scaling form:

\begin{enumerate}[(1)]
  \item Note that for the usual order parameter $M^2$, the case that $M^2$ is close to its saturate value usually happens beyond the critical region. But for the angular order parameter $\phi_q$, $\phi_q$ can approach its saturate value $1$ in the critical region for small $g$ and large $L$. Thus, the scaling behaviors for different ranges of $gL^{1/\nu}$ and $gL^{1/\nu'}$ are generally covered in Eq.~(\ref{eq:equiphi1}).

  \item The necessity of an additional $\nu'$ which is different from $\nu$ can be seen from the two limits that $\phi_q$ is dimensionless when $\phi_q$ is large (Eq.~(\ref{eq:equiphi4})) and $\phi_q\propto L^{-|y_q|}$ when $\phi_q$ is small (Eq.~(\ref{eq:equiphi2})). To see this, we assume the exponent in Eq.~(\ref{eq:equiphi4}) (for large $\phi_q$) were still $\nu$, we find that Eq.~(\ref{eq:equiphi3}) cannot be analytically converted to Eq.~(\ref{eq:equiphi4}), as the term of $L^{-|y_q|}$ cannot disappear into thin air. Thus $\nu'$ must be different from $\nu$, since the information of $y_q$ must be reflected in $\nu'$.

  \item These scaling analyses are consistent with the numerical results of the Monte Carlo renormalization group analyses, as discussed in Ref.~\cite{Shao2020prl}.

  \item The scaling law connecting $\nu'$ and $\nu$ obtained in Ref.~\cite{Shao2020prl} is also consistent with that obtained via functional renormalization group analyses~\cite{Leonard2015prl}.
\end{enumerate}

\subsubsection{\label{dy}Dynamic scaling forms}
When considering the nonequilibrium relaxation dynamics, one may naturally generalize the scaling form as
\begin{equation}
  \phi_q(g,L,t)=L^{-|y_q|}f_{\phi1}(gL^{1/\nu},gL^{1/\nu'},tL^{-z}).
  \label{eq:equidyphi1}
\end{equation}
However, our numerical results show that Eq.~(\ref{eq:equidyphi1}) with a single dynamic exponent $z$ is inadequate to cover all universal properties. Rather, another dynamic exponent $z'$ should be included in the scaling form, which becomes
\begin{equation}
  \phi_q(g,L,t)=L^{-|y_q|}f_{\phi2}(gL^{1/\nu},gL^{1/\nu'},tL^{-z},tL^{-z'}).
  \label{eq:equidyphi2}
\end{equation}
Apparently, the appearance of $z'$ must be related to DISV. In the nonequilibrium process, the critical slowing down causes some universal information stored in the microscopic Hamiltonian and the initial state to be remembered. In this way, there can be an additional scaling variable $\mathcal{G}(t,\lambda_q,g_0)$, which is similar to $\mathcal{F}_q(g,\lambda_q)$ and is a composite operator of time $t$, $\lambda_q$ in the microscopic Hamiltonian, and $g_0$ the initial state information. Consequently, the full scaling form reads
\begin{equation}
  Q(g,\lambda_q,k,t)=k^{\kappa}f_{Q4}[kg^{-\nu},k\mathcal{F}_q(g,\lambda_q),kt^{1/z},k\mathcal{G}(t,\lambda_q,g_0))].
\end{equation}
Although the detailed form of $\mathcal{G}(t,\lambda_q,g_0)$ is unknown from pure scaling analyses, one expects that this term can contribute a new dynamic exponent $z'$, similar to the case that $\mathcal{F}_q(g,\lambda_q)$ gives a new correlation length exponent $\nu'$. Moreover, $z'$ should be related to the transverse mass. This can be seen in Fig. 2 in the main text.

\begin{figure*}[!htbp]
\centering
  \includegraphics[width=\linewidth,clip]{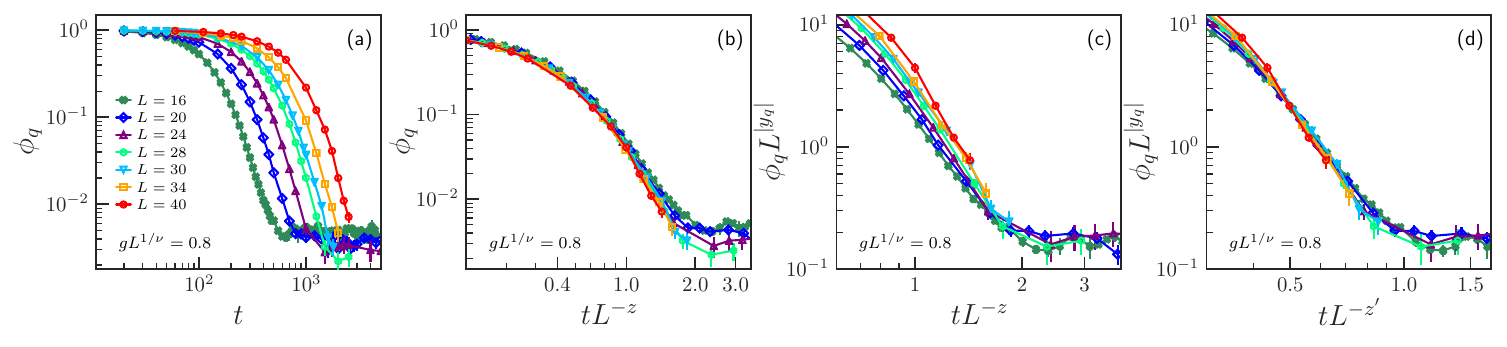}
  \vskip-3mm
  \caption{Off-critical-point effect for $g>0$ with fixed $gL^{1/\nu}=0.8$ for systems with different size with $q=5$. (a) Evolution of $\phi_q$ with $t$. (b) At early times, $\phi_q$ is dimensionless and governed by $z$. (c) At long times, $\phi_q$ has a dimension $-y_q$. However, with $t$ rescaled as $tL^{-z}$, the rescaled curves cannot match. (d) Instead, when $t$ is rescaled as $tL^{-z'}$ with $z'=2.25$, good collapse of the rescaled curves is observed.
}
\label{fig:gLp}
\end{figure*}

Figures.~2 and Fig. 3 in the main text show the scaling behaviors of $\phi_q$ from the ordered initial state in which $\phi_q=1$. In the initial stage, $\phi_q$ is close to $1$. Thus, it should be dimensionless as discussed above (See also Ref.~\cite{Shao2020prl}). In this short-time stage, the scaling form is given by
\begin{equation}
 \phi_q(t)=f_{\phi3}(tL^{-z}).
 \label{eq:equidyphi3}
\end{equation}
In the long-time stage, $\phi_q\propto L^{-|y_q|}$. The scaling form in the long-time stage becomes
\begin{equation}
 \phi_q(t)=L^{-|y_q|}f_{\phi4}(tL^{-z'}).
 \label{eq:equidyphi5}
\end{equation}
These scaling analyses are consistent with the numerical results shown in Figs.~2 and 3.

The dynamic scaling forms of Eqs.~(\ref{eq:equidyphi3}) and (\ref{eq:equidyphi5}) are very similar to the equilibrium scaling forms of Eqs.~(\ref{eq:equiphi2}) and (\ref{eq:equiphi4}) in the sense that the former connect the cases of large $\phi_q$ (dimensionless) and small $\phi_q$ (with scaling dimension $|y_q|$) in time direction; while the latter connect the cases of large $\phi_q$ (dimensionless) and small $\phi_q$ (with scaling dimension $|y_q|$) in off-critical-point temperature direction. Thus, the appearance of $z'$ is similar to the appearance of $\nu'$ and both  are induced by the scaling with DISV.

Moreover, scaling analyses similar to Remarks (1) and (2) above Eq.~(\ref{eq:emefull4}) demonstrate that $z'$ must be introduced as an indispensable critical exponent. If the dynamic exponent in the long-time stage were still $z$, the scaling form becomes
\begin{equation}
 \phi_q(t)=L^{-|y_q|}f_{\phi5}(tL^{-z}).
 \label{eq:equidyphi4}
\end{equation}
However, the scaling form Eq.~(\ref{eq:equidyphi4}) cannot be generated from Eq.~(\ref{eq:equidyphi3}) analytically, since the critical exponent $y_q$ cannot come from thin air. Consequently, there must exist a new dynamic exponent $z'$, combining $z$ and $y_0$ (although the precise relation is unknown), similar to the case that $\nu'$ combines $\nu$ and $y_0$, which is just the scaling form in Eq.~(\ref{eq:equidyphi5}).

\begin{figure*}[!htbp]
\centering
  \includegraphics[width=\linewidth,clip]{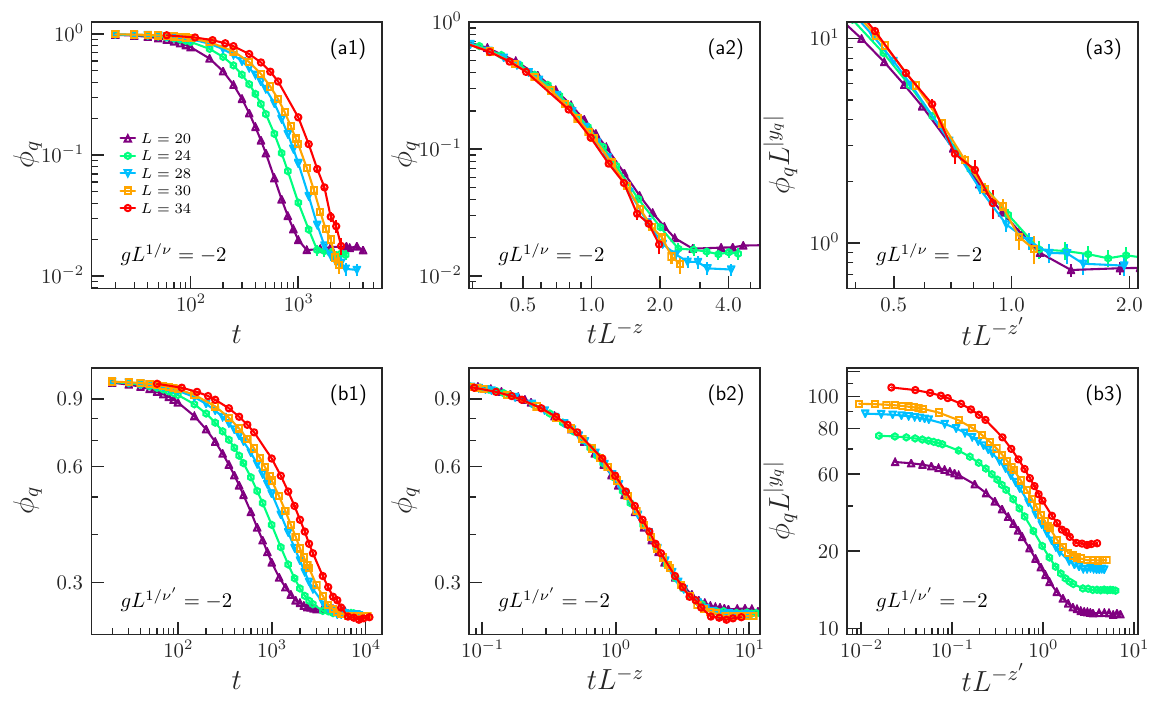}
  \vskip-3mm
  \caption{Off-critical-point effect for $g<0$ with fixed $gL^{1/\nu}=-2$ and $gL^{1/\nu'}=-2$.
    (a1)-(a3): Results of fixed $gL^{1/\nu}=-2$. $\phi_q$ exhibits a two-stage evolution, similar to the case of $gL^{1/\nu}=-0.8$ shown in the main text.
    (b1)-(b3): Results of fixed $gL^{1/\nu'}=-2$. The behaviors of $\phi_q$ differ from those in (a1)-(a3). $\phi_q$ is dimensionless throughout the entire relaxation and the governing dynamic exponent is $z$. The good collapses in both (a2) and (b2) suggest a complex crossover regime where both $\xi$ and $\xi'$ play roles.
  }
  \label{fig:compgg}
\end{figure*}

\begin{figure*}[!htbp]
\centering
  \includegraphics[width=\linewidth,clip]{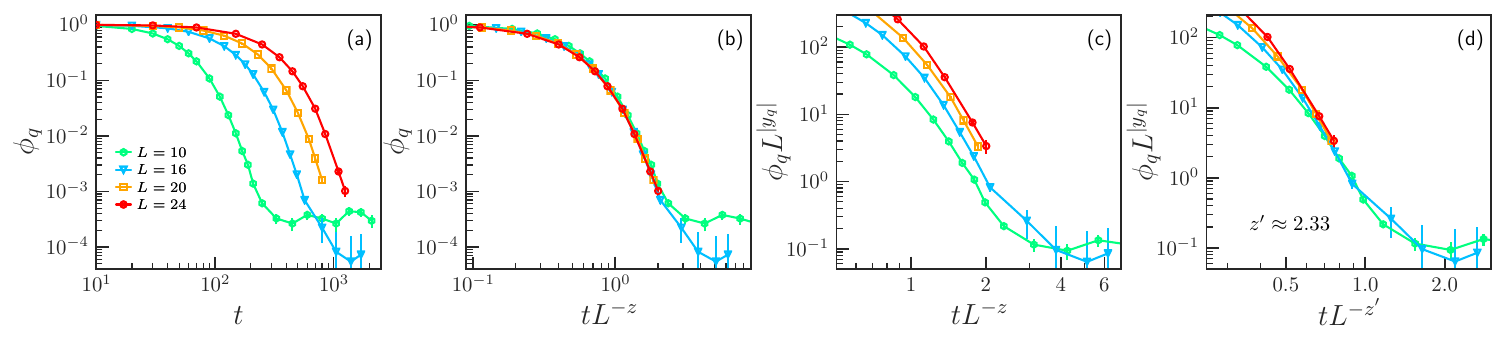}
  \vskip-3mm
  \caption{Relaxation dynamics of $\phi_q$ for the anisotropic clock model with $q=6$ and an anisotropy strength $\lambda=0.5$. (a) Time-evolution of $\phi_q$ for different system sizes. (b) Similar to the isotropic case, in the early stage, $\phi_q$ is dimensionless and controlled the usual $z$. (c) At long times, $\phi_q$ is rescaled as $\phi_qL^{|y_q|}$ and $t$ as $tL^{-z}$, but the usual $z$ is not able to capture the dynamics. (d) Instead, rescaling $t$ as $tL^{-z'}$ with $z'=2.33$, the long-time behavior of $\phi_q$ is well described.}
\label{fig:aniq6}
\end{figure*}

\section{\label{Sec2}Off-critical effects for $g>0$}

In the main text, we show that with $gL^{1/\nu}=-0.8$ for $q=5$, the scaling behaviors of $\phi_q$ are consistent with those at the critical point. Here we consider the case of $gL^{1/\nu}=0.8$. As shown in Fig.~\ref{fig:gLp}(a)-(b), similar to the case of $gL^{1/\nu}=-0.8$, the evolution of $\phi_q$ also displays a two-stage behavior. During short-time stage, $\phi_q$ is dimensionless and characterized by the dynamic exponent $z$ belonging to the XY universality class. In the long-time stage, $\phi_q$ has a dimension $-y_q$ and is controlled by $z'$ with the same value $z'=2.25$.

These findings demonstrate that the scaling form $\phi_q$ also holds in the  $g>0$ off-critical region when $|g|L^{1/\nu}$ is small, with $z'$ maintaining the same value with the $g=0$ case.

\section{\label{Sec3}Crossover behaviors for larger $g<0$}

On the ordered side, namely when $g<0$, the equilibrium critical properties of $\phi_q$ have attracted long-term investigations. It was shown by Ref.~\cite{Shao2020prl} that as $gL^{1/\nu}$ and $gL^{1/\nu'}$ increase, $gL^{1/\nu'}$ becomes the dominated variable. Here we explore the scaling of $\phi_q$ for a larger fixed value of the scaling variables $gL^{1/\nu}$ and $gL^{1/\nu'}$ when $g<0$.

As discussed in the main text and the previous sections, the full scaling form of $\phi_q$ should follows
\begin{equation}
  \phi_q(g,L,t)=L^{-\kappa}f_{\phi}(tL^{-z},tL^{-z'};gL^{1/\nu},gL^{1/\nu'}),
  \label{eq:full}
\end{equation}
in which the scaling dimension $\kappa=0$ for the short-time stage and $\kappa=-y_q$ for the long-time stage. Moving away from the critical point, we need to account for the effects of the two scaling variables $gL^{1/\nu}$ and $gL^{1/\nu'}$. As demonstrated in the main text, for $g<0$, when $|g|L^{1/\nu}$ is small and $|g|L^{1/\nu'}$ is even smaller, $gL^{1/\nu}$ dominates the behavior of $f_{\phi}$.

Here we focus on larger $|g|L^{1/\nu}$ and $|g|L^{1/\nu'}$ for $g<0$. In Fig.~\ref{fig:compgg}(a1)-(a3), with fixed $gL^{1/\nu}=-2$, the dynamics of $\phi_q$ also exhibits a two-stage evolution, governed by $z$ and $z'$ for short times and long times, respectively. In the short-time stage, $\phi_q$ satisfies
\begin{equation}
  \phi_q(g,L,t)=f_{\phi}(tL^{-z},gL^{1/\nu}),
  \label{eq:full1}
\end{equation}
whereas in the long-time stage, $\phi_q(g,L,t)$ obeys
\begin{equation}
  \phi_q(g,L,t)=L^{-|y_q|}f_{\phi}(tL^{-z'},gL^{1/\nu}).
  \label{eq:full2}
\end{equation}
These results are consistent with those presented in the main text with $gL^{1/\nu}=-0.8$, demonstrating that the scaling form is universal.

Moreover, for fixed $gL^{1/\nu'}=-2$, as shown in Fig.~\ref{fig:compgg}(b1)-(b2), in the whole stage, $\phi_q$ satisfies
\begin{equation}
  \phi_q(g,L,t)=f_{\phi}(tL^{-z},gL^{1/\nu'}).
  \label{eq:full3}
\end{equation}
Such behaviors indicate that $\phi_q$ no longer has a dimension, as confirmed in Fig.~\ref{fig:compgg}(b3). When $\phi_q$ is rescaled as $\phi_q L^{|y_q|}$, the rescaled curves cannot match. In the thermodynamic limit, Eq.~(\ref{eq:full3}) tends to $\phi_q(g,L,t)=f_{\phi}(gL^{1/\nu'})$, which is consistent with the equilibrium scaling form for larger $\phi_q$~\cite{Shao2020prl}. These results demonstrate that for larger $gL^{1/\nu}$, $z'$ fades out in the long-time scaling behaviors. The reason is that for large $gL^{1/\nu'}$, the transverse fluctuations are weak. Thus, these results confirm from the reverse side that $z'$ plays roles when the transverse fluctuations dominate, as discussed in the main text.

\section{Anisotropic clock model}

To verify that the scaling form of $\phi_q$ is universal, we also consider an anisotropic clock model on a cubic lattice~\cite{Patil2021prb}

\begin{equation}
  H=-J_{\parallel}\sum_{\langle i,j\rangle_{\parallel}}\cos(\theta_i-\theta_j)-J_{\perp}\sum_{\langle i,j\rangle_{\perp}}\cos(\theta_i-\theta_j),
\end{equation}
in which $\parallel$ and $\perp$ denotes nearest-neighbor pairs $\langle i,j\rangle$ in the $xy$ plane and $z$ direction. For simplicity, we define the couplings using the anisotropy parameter $\lambda\in [0,1)$ with $J_{\parallel}=(1-\lambda)$ and $J_{\perp}=(1+\lambda)$. The anisotropy is anticipated to be relevant in resulting a crossover behavior in the value of $\nu'$~\cite{Patil2021prb} but its impact on the relaxation dynamics remains unknown.

Here we consider the relaxation dynamics for the $q=6$ model with $\lambda=0.5$ at the critical point $T_c=1.697(1)$~\cite{Patil2021prb}. Similarly, the two-stage character of $\phi_q$ is also observed in Fig.~\ref{fig:aniq6} during the evolution started from an ordered initial state. At short times, $\phi_q$ is dimensionless and governed by $z$, while at long times, $\phi_q$ should be rescaled as $\phi_q L^{|y_q|}$ with $y_q=-2.55$~\cite{Shao2020prl} and controlled by $z'=2.33$ with the same value as the isotropic case with $q=6$.

These results soundly support that the scaling form of $\phi_q$ is universal and the value of $z'$ is independent on the strength of anisotropy. Thus, $z'$ is a crucial exponent in characterizing the relaxation dynamics involving the dangerously irrelevant scaling variable.

\bibliography{refzq}

\begin{thebibliography}{65}%
\makeatletter
\providecommand \@ifxundefined [1]{%
 \@ifx{#1\undefined}
}%
\providecommand \@ifnum [1]{%
 \ifnum #1\expandafter \@firstoftwo
 \else \expandafter \@secondoftwo
 \fi
}%
\providecommand \@ifx [1]{%
 \ifx #1\expandafter \@firstoftwo
 \else \expandafter \@secondoftwo
 \fi
}%
\providecommand \natexlab [1]{#1}%
\providecommand \enquote  [1]{``#1''}%
\providecommand \bibnamefont  [1]{#1}%
\providecommand \bibfnamefont [1]{#1}%
\providecommand \citenamefont [1]{#1}%
\providecommand \href@noop [0]{\@secondoftwo}%
\providecommand \href [0]{\begingroup \@sanitize@url \@href}%
\providecommand \@href[1]{\@@startlink{#1}\@@href}%
\providecommand \@@href[1]{\endgroup#1\@@endlink}%
\providecommand \@sanitize@url [0]{\catcode `\\12\catcode `\$12\catcode
  `\&12\catcode `\#12\catcode `\^12\catcode `\_12\catcode `\%12\relax}%
\providecommand \@@startlink[1]{}%
\providecommand \@@endlink[0]{}%
\providecommand \url  [0]{\begingroup\@sanitize@url \@url }%
\providecommand \@url [1]{\endgroup\@href {#1}{\urlprefix }}%
\providecommand \urlprefix  [0]{URL }%
\providecommand \Eprint [0]{\href }%
\providecommand \doibase [0]{http://dx.doi.org/}%
\providecommand \selectlanguage [0]{\@gobble}%
\providecommand \bibinfo  [0]{\@secondoftwo}%
\providecommand \bibfield  [0]{\@secondoftwo}%
\providecommand \translation [1]{[#1]}%
\providecommand \BibitemOpen [0]{}%
\providecommand \bibitemStop [0]{}%
\providecommand \bibitemNoStop [0]{.\EOS\space}%
\providecommand \EOS [0]{\spacefactor3000\relax}%
\providecommand \BibitemShut  [1]{\csname bibitem#1\endcsname}%
\let\auto@bib@innerbib\@empty
\bibitem [{\citenamefont {Landau}\ \emph {et~al.}(1999)\citenamefont {Landau},
  \citenamefont {Lifshitz},\ and\ \citenamefont {Pitaevskii}}]{Landaubook}%
  \BibitemOpen
  \bibfield  {author} {\bibinfo {author} {\bibfnamefont {L.~D.}\ \bibnamefont
  {Landau}}, \bibinfo {author} {\bibfnamefont {E.~M.}\ \bibnamefont
  {Lifshitz}}, \ and\ \bibinfo {author} {\bibfnamefont {E.~M.}\ \bibnamefont
  {Pitaevskii}},\ }\href@noop {} {\emph {\bibinfo {title} {Statistical
  Physics}}}\ (\bibinfo  {publisher} {Butterworth-Heinemann},\ \bibinfo
  {address} {New York},\ \bibinfo {year} {1999})\BibitemShut {NoStop}%
\bibitem [{\citenamefont {Oshikawa}(2000)}]{Oshikawa2000prb}%
  \BibitemOpen
  \bibfield  {author} {\bibinfo {author} {\bibfnamefont {M.}~\bibnamefont
  {Oshikawa}},\ }\bibfield  {title} {\enquote {\bibinfo {title} {{Ordered phase
  and scaling in ${Z}_{n}$ models and the three-state antiferromagnetic Potts
  model in three dimensions}},}\ }\href {\doibase 10.1103/PhysRevB.61.3430}
  {\bibfield  {journal} {\bibinfo  {journal} {Phys. Rev. B}\ }\textbf {\bibinfo
  {volume} {61}},\ \bibinfo {pages} {3430--3434} (\bibinfo {year}
  {2000})}\BibitemShut {NoStop}%
\bibitem [{\citenamefont {Lou}\ \emph {et~al.}(2007)\citenamefont {Lou},
  \citenamefont {Sandvik},\ and\ \citenamefont {Balents}}]{Lou2007prl}%
  \BibitemOpen
  \bibfield  {author} {\bibinfo {author} {\bibfnamefont {J.}~\bibnamefont
  {Lou}}, \bibinfo {author} {\bibfnamefont {A.~W.}\ \bibnamefont {Sandvik}}, \
  and\ \bibinfo {author} {\bibfnamefont {L.}~\bibnamefont {Balents}},\
  }\bibfield  {title} {\enquote {\bibinfo {title} {{Emergence of U(1) Symmetry
  in the 3D $XY$ Model with ${Z}_{q}$ Anisotropy}},}\ }\href {\doibase
  10.1103/PhysRevLett.99.207203} {\bibfield  {journal} {\bibinfo  {journal}
  {Phys. Rev. Lett.}\ }\textbf {\bibinfo {volume} {99}},\ \bibinfo {pages}
  {207203} (\bibinfo {year} {2007})}\BibitemShut {NoStop}%
\bibitem [{\citenamefont {Okubo}\ \emph {et~al.}(2015)\citenamefont {Okubo},
  \citenamefont {Oshikawa}, \citenamefont {Watanabe},\ and\ \citenamefont
  {Kawashima}}]{Okubo2015prb}%
  \BibitemOpen
  \bibfield  {author} {\bibinfo {author} {\bibfnamefont {T.}~\bibnamefont
  {Okubo}}, \bibinfo {author} {\bibfnamefont {K.}~\bibnamefont {Oshikawa}},
  \bibinfo {author} {\bibfnamefont {H.}~\bibnamefont {Watanabe}}, \ and\
  \bibinfo {author} {\bibfnamefont {N.}~\bibnamefont {Kawashima}},\ }\bibfield
  {title} {\enquote {\bibinfo {title} {{Scaling relation for dangerously
  irrelevant symmetry-breaking fields}},}\ }\href {\doibase
  10.1103/PhysRevB.91.174417} {\bibfield  {journal} {\bibinfo  {journal} {Phys.
  Rev. B}\ }\textbf {\bibinfo {volume} {91}},\ \bibinfo {pages} {174417}
  (\bibinfo {year} {2015})}\BibitemShut {NoStop}%
\bibitem [{\citenamefont {L\'eonard}\ and\ \citenamefont
  {Delamotte}(2015)}]{Leonard2015prl}%
  \BibitemOpen
  \bibfield  {author} {\bibinfo {author} {\bibfnamefont {F.}~\bibnamefont
  {L\'eonard}}\ and\ \bibinfo {author} {\bibfnamefont {B.}~\bibnamefont
  {Delamotte}},\ }\bibfield  {title} {\enquote {\bibinfo {title} {{Critical
  Exponents Can Be Different on the Two Sides of a Transition: A Generic
  Mechanism}},}\ }\href {\doibase 10.1103/PhysRevLett.115.200601} {\bibfield
  {journal} {\bibinfo  {journal} {Phys. Rev. Lett.}\ }\textbf {\bibinfo
  {volume} {115}},\ \bibinfo {pages} {200601} (\bibinfo {year}
  {2015})}\BibitemShut {NoStop}%
\bibitem [{\citenamefont {Pujari}\ \emph {et~al.}(2015)\citenamefont {Pujari},
  \citenamefont {Alet},\ and\ \citenamefont {Damle}}]{Pujari2015prb}%
  \BibitemOpen
  \bibfield  {author} {\bibinfo {author} {\bibfnamefont {S.}~\bibnamefont
  {Pujari}}, \bibinfo {author} {\bibfnamefont {F.}~\bibnamefont {Alet}}, \ and\
  \bibinfo {author} {\bibfnamefont {K.}~\bibnamefont {Damle}},\ }\bibfield
  {title} {\enquote {\bibinfo {title} {{Transitions to valence-bond solid order
  in a honeycomb lattice antiferromagnet}},}\ }\href {\doibase
  10.1103/PhysRevB.91.104411} {\bibfield  {journal} {\bibinfo  {journal} {Phys.
  Rev. B}\ }\textbf {\bibinfo {volume} {91}},\ \bibinfo {pages} {104411}
  (\bibinfo {year} {2015})}\BibitemShut {NoStop}%
\bibitem [{\citenamefont {Ding}\ \emph {et~al.}(2016)\citenamefont {Ding},
  \citenamefont {Bl\"ote},\ and\ \citenamefont {Deng}}]{Ding2016prb}%
  \BibitemOpen
  \bibfield  {author} {\bibinfo {author} {\bibfnamefont {C.}~\bibnamefont
  {Ding}}, \bibinfo {author} {\bibfnamefont {H.~W.~J.}\ \bibnamefont
  {Bl\"ote}}, \ and\ \bibinfo {author} {\bibfnamefont {Y.}~\bibnamefont
  {Deng}},\ }\bibfield  {title} {\enquote {\bibinfo {title} {{Emergent O($n$)
  symmetry in a series of three-dimensional Potts models}},}\ }\href {\doibase
  10.1103/PhysRevB.94.104402} {\bibfield  {journal} {\bibinfo  {journal} {Phys.
  Rev. B}\ }\textbf {\bibinfo {volume} {94}},\ \bibinfo {pages} {104402}
  (\bibinfo {year} {2016})}\BibitemShut {NoStop}%
\bibitem [{\citenamefont {Hasenbusch}\ and\ \citenamefont
  {Vicari}(2011)}]{Hasenbusch2011prb}%
  \BibitemOpen
  \bibfield  {author} {\bibinfo {author} {\bibfnamefont {M.}~\bibnamefont
  {Hasenbusch}}\ and\ \bibinfo {author} {\bibfnamefont {E.}~\bibnamefont
  {Vicari}},\ }\bibfield  {title} {\enquote {\bibinfo {title} {{Anisotropic
  perturbations in three-dimensional O($N$)-symmetric vector models}},}\ }\href
  {\doibase 10.1103/PhysRevB.84.125136} {\bibfield  {journal} {\bibinfo
  {journal} {Phys. Rev. B}\ }\textbf {\bibinfo {volume} {84}},\ \bibinfo
  {pages} {125136} (\bibinfo {year} {2011})}\BibitemShut {NoStop}%
\bibitem [{\citenamefont {Shao}\ \emph {et~al.}(2020)\citenamefont {Shao},
  \citenamefont {Guo},\ and\ \citenamefont {Sandvik}}]{Shao2020prl}%
  \BibitemOpen
  \bibfield  {author} {\bibinfo {author} {\bibfnamefont {H.}~\bibnamefont
  {Shao}}, \bibinfo {author} {\bibfnamefont {W.}~\bibnamefont {Guo}}, \ and\
  \bibinfo {author} {\bibfnamefont {A.~W.}\ \bibnamefont {Sandvik}},\
  }\bibfield  {title} {\enquote {\bibinfo {title} {{Monte Carlo Renormalization
  Flows in the Space of Relevant and Irrelevant Operators: Application to
  Three-Dimensional Clock Models}},}\ }\href {\doibase
  10.1103/PhysRevLett.124.080602} {\bibfield  {journal} {\bibinfo  {journal}
  {Phys. Rev. Lett.}\ }\textbf {\bibinfo {volume} {124}},\ \bibinfo {pages}
  {080602} (\bibinfo {year} {2020})}\BibitemShut {NoStop}%
\bibitem [{\citenamefont {Patil}\ \emph {et~al.}(2021)\citenamefont {Patil},
  \citenamefont {Shao},\ and\ \citenamefont {Sandvik}}]{Patil2021prb}%
  \BibitemOpen
  \bibfield  {author} {\bibinfo {author} {\bibfnamefont {P.}~\bibnamefont
  {Patil}}, \bibinfo {author} {\bibfnamefont {H.}~\bibnamefont {Shao}}, \ and\
  \bibinfo {author} {\bibfnamefont {A.~W.}\ \bibnamefont {Sandvik}},\
  }\bibfield  {title} {\enquote {\bibinfo {title} {{Unconventional U(1) to
  ${Z}_{q}$ crossover in quantum and classical $q$-state clock models}},}\
  }\href {\doibase 10.1103/PhysRevB.103.054418} {\bibfield  {journal} {\bibinfo
   {journal} {Phys. Rev. B}\ }\textbf {\bibinfo {volume} {103}},\ \bibinfo
  {pages} {054418} (\bibinfo {year} {2021})}\BibitemShut {NoStop}%
\bibitem [{\citenamefont {Li}\ \emph {et~al.}(2017)\citenamefont {Li},
  \citenamefont {Jiang}, \citenamefont {Jian},\ and\ \citenamefont
  {Yao}}]{Li2017nc}%
  \BibitemOpen
  \bibfield  {author} {\bibinfo {author} {\bibfnamefont {Z.-X.}\ \bibnamefont
  {Li}}, \bibinfo {author} {\bibfnamefont {Y.-F.}\ \bibnamefont {Jiang}},
  \bibinfo {author} {\bibfnamefont {S.-K.}\ \bibnamefont {Jian}}, \ and\
  \bibinfo {author} {\bibfnamefont {H.}~\bibnamefont {Yao}},\ }\bibfield
  {title} {\enquote {\bibinfo {title} {{Fermion-induced quantum critical
  points}},}\ }\href {\doibase 10.1038/s41467-017-00167-6} {\bibfield
  {journal} {\bibinfo  {journal} {Nature Communications}\ }\textbf {\bibinfo
  {volume} {8}},\ \bibinfo {pages} {314} (\bibinfo {year} {2017})}\BibitemShut
  {NoStop}%
\bibitem [{\citenamefont {Jian}\ and\ \citenamefont
  {Yao}(2017{\natexlab{a}})}]{Jian2017prb}%
  \BibitemOpen
  \bibfield  {author} {\bibinfo {author} {\bibfnamefont {S.-K.}\ \bibnamefont
  {Jian}}\ and\ \bibinfo {author} {\bibfnamefont {H.}~\bibnamefont {Yao}},\
  }\bibfield  {title} {\enquote {\bibinfo {title} {Fermion-induced quantum
  critical points in three-dimensional weyl semimetals},}\ }\href {\doibase
  10.1103/PhysRevB.96.155112} {\bibfield  {journal} {\bibinfo  {journal} {Phys.
  Rev. B}\ }\textbf {\bibinfo {volume} {96}},\ \bibinfo {pages} {155112}
  (\bibinfo {year} {2017}{\natexlab{a}})}\BibitemShut {NoStop}%
\bibitem [{\citenamefont {Jian}\ and\ \citenamefont
  {Yao}(2017{\natexlab{b}})}]{Jian2018prb}%
  \BibitemOpen
  \bibfield  {author} {\bibinfo {author} {\bibfnamefont {S.-K.}\ \bibnamefont
  {Jian}}\ and\ \bibinfo {author} {\bibfnamefont {H.}~\bibnamefont {Yao}},\
  }\bibfield  {title} {\enquote {\bibinfo {title} {Fermion-induced quantum
  critical points in two-dimensional dirac semimetals},}\ }\href {\doibase
  10.1103/PhysRevB.96.195162} {\bibfield  {journal} {\bibinfo  {journal} {Phys.
  Rev. B}\ }\textbf {\bibinfo {volume} {96}},\ \bibinfo {pages} {195162}
  (\bibinfo {year} {2017}{\natexlab{b}})}\BibitemShut {NoStop}%
\bibitem [{\citenamefont {Torres}\ \emph {et~al.}(2018)\citenamefont {Torres},
  \citenamefont {Classen}, \citenamefont {Herbut},\ and\ \citenamefont
  {Scherer}}]{Torres2018prb}%
  \BibitemOpen
  \bibfield  {author} {\bibinfo {author} {\bibfnamefont {E.}~\bibnamefont
  {Torres}}, \bibinfo {author} {\bibfnamefont {L.}~\bibnamefont {Classen}},
  \bibinfo {author} {\bibfnamefont {I.~F.}\ \bibnamefont {Herbut}}, \ and\
  \bibinfo {author} {\bibfnamefont {M.~M.}\ \bibnamefont {Scherer}},\
  }\bibfield  {title} {\enquote {\bibinfo {title} {Fermion-induced quantum
  criticality with two length scales in dirac systems},}\ }\href {\doibase
  10.1103/PhysRevB.97.125137} {\bibfield  {journal} {\bibinfo  {journal} {Phys.
  Rev. B}\ }\textbf {\bibinfo {volume} {97}},\ \bibinfo {pages} {125137}
  (\bibinfo {year} {2018})}\BibitemShut {NoStop}%
\bibitem [{\citenamefont {Classen}\ \emph {et~al.}(2017)\citenamefont
  {Classen}, \citenamefont {Herbut},\ and\ \citenamefont
  {Scherer}}]{Classen2017prb}%
  \BibitemOpen
  \bibfield  {author} {\bibinfo {author} {\bibfnamefont {L.}~\bibnamefont
  {Classen}}, \bibinfo {author} {\bibfnamefont {I.~F.}\ \bibnamefont {Herbut}},
  \ and\ \bibinfo {author} {\bibfnamefont {M.~M.}\ \bibnamefont {Scherer}},\
  }\bibfield  {title} {\enquote {\bibinfo {title} {Fluctuation-induced
  continuous transition and quantum criticality in dirac semimetals},}\ }\href
  {\doibase 10.1103/PhysRevB.96.115132} {\bibfield  {journal} {\bibinfo
  {journal} {Phys. Rev. B}\ }\textbf {\bibinfo {volume} {96}},\ \bibinfo
  {pages} {115132} (\bibinfo {year} {2017})}\BibitemShut {NoStop}%
\bibitem [{\citenamefont {Senthil}\ \emph
  {et~al.}(2004{\natexlab{a}})\citenamefont {Senthil}, \citenamefont
  {Vishwanath}, \citenamefont {Balents}, \citenamefont {Sachdev},\ and\
  \citenamefont {Fisher}}]{Senthil2004sci}%
  \BibitemOpen
  \bibfield  {author} {\bibinfo {author} {\bibfnamefont {T.}~\bibnamefont
  {Senthil}}, \bibinfo {author} {\bibfnamefont {A.}~\bibnamefont {Vishwanath}},
  \bibinfo {author} {\bibfnamefont {L.}~\bibnamefont {Balents}}, \bibinfo
  {author} {\bibfnamefont {S.}~\bibnamefont {Sachdev}}, \ and\ \bibinfo
  {author} {\bibfnamefont {M.~P.~A.}\ \bibnamefont {Fisher}},\ }\bibfield
  {title} {\enquote {\bibinfo {title} {Deconfined quantum critical points},}\
  }\href {\doibase 10.1126/science.1091806} {\bibfield  {journal} {\bibinfo
  {journal} {Science}\ }\textbf {\bibinfo {volume} {303}},\ \bibinfo {pages}
  {1490--1494} (\bibinfo {year} {2004}{\natexlab{a}})}\BibitemShut {NoStop}%
\bibitem [{\citenamefont {Senthil}\ \emph
  {et~al.}(2004{\natexlab{b}})\citenamefont {Senthil}, \citenamefont {Balents},
  \citenamefont {Sachdev}, \citenamefont {Vishwanath},\ and\ \citenamefont
  {Fisher}}]{Senthil2004prb}%
  \BibitemOpen
  \bibfield  {author} {\bibinfo {author} {\bibfnamefont {T.}~\bibnamefont
  {Senthil}}, \bibinfo {author} {\bibfnamefont {L.}~\bibnamefont {Balents}},
  \bibinfo {author} {\bibfnamefont {S.}~\bibnamefont {Sachdev}}, \bibinfo
  {author} {\bibfnamefont {A.}~\bibnamefont {Vishwanath}}, \ and\ \bibinfo
  {author} {\bibfnamefont {M.~P.~A.}\ \bibnamefont {Fisher}},\ }\bibfield
  {title} {\enquote {\bibinfo {title} {{Quantum criticality beyond the
  Landau-Ginzburg-Wilson paradigm}},}\ }\href {\doibase
  10.1103/PhysRevB.70.144407} {\bibfield  {journal} {\bibinfo  {journal} {Phys.
  Rev. B}\ }\textbf {\bibinfo {volume} {70}},\ \bibinfo {pages} {144407}
  (\bibinfo {year} {2004}{\natexlab{b}})}\BibitemShut {NoStop}%
\bibitem [{\citenamefont {Nahum}\ \emph {et~al.}(2015)\citenamefont {Nahum},
  \citenamefont {Serna}, \citenamefont {Chalker}, \citenamefont {Ortu\~no},\
  and\ \citenamefont {Somoza}}]{Nahum2015prl}%
  \BibitemOpen
  \bibfield  {author} {\bibinfo {author} {\bibfnamefont {A.}~\bibnamefont
  {Nahum}}, \bibinfo {author} {\bibfnamefont {P.}~\bibnamefont {Serna}},
  \bibinfo {author} {\bibfnamefont {J.~T.}\ \bibnamefont {Chalker}}, \bibinfo
  {author} {\bibfnamefont {M.}~\bibnamefont {Ortu\~no}}, \ and\ \bibinfo
  {author} {\bibfnamefont {A.~M.}\ \bibnamefont {Somoza}},\ }\bibfield  {title}
  {\enquote {\bibinfo {title} {{Emergent SO($5$) Symmetry at the N\'eel to
  Valence-Bond-Solid Transition}},}\ }\href {\doibase
  10.1103/PhysRevLett.115.267203} {\bibfield  {journal} {\bibinfo  {journal}
  {Phys. Rev. Lett.}\ }\textbf {\bibinfo {volume} {115}},\ \bibinfo {pages}
  {267203} (\bibinfo {year} {2015})}\BibitemShut {NoStop}%
\bibitem [{\citenamefont {Wang}\ \emph {et~al.}(2017)\citenamefont {Wang},
  \citenamefont {Nahum}, \citenamefont {Metlitski}, \citenamefont {Xu},\ and\
  \citenamefont {Senthil}}]{Wang2017prx}%
  \BibitemOpen
  \bibfield  {author} {\bibinfo {author} {\bibfnamefont {C.}~\bibnamefont
  {Wang}}, \bibinfo {author} {\bibfnamefont {A.}~\bibnamefont {Nahum}},
  \bibinfo {author} {\bibfnamefont {M.~A.}\ \bibnamefont {Metlitski}}, \bibinfo
  {author} {\bibfnamefont {C.}~\bibnamefont {Xu}}, \ and\ \bibinfo {author}
  {\bibfnamefont {T.}~\bibnamefont {Senthil}},\ }\bibfield  {title} {\enquote
  {\bibinfo {title} {{Deconfined Quantum Critical Points: Symmetries and
  Dualities}},}\ }\href {\doibase 10.1103/PhysRevX.7.031051} {\bibfield
  {journal} {\bibinfo  {journal} {Phys. Rev. X}\ }\textbf {\bibinfo {volume}
  {7}},\ \bibinfo {pages} {031051} (\bibinfo {year} {2017})}\BibitemShut
  {NoStop}%
\bibitem [{\citenamefont {Takahashi}\ and\ \citenamefont
  {Sandvik}(2020)}]{Takahashi2020prr}%
  \BibitemOpen
  \bibfield  {author} {\bibinfo {author} {\bibfnamefont {J.}~\bibnamefont
  {Takahashi}}\ and\ \bibinfo {author} {\bibfnamefont {A.~W.}\ \bibnamefont
  {Sandvik}},\ }\bibfield  {title} {\enquote {\bibinfo {title} {{Valence-bond
  solids, vestigial order, and emergent SO(5) symmetry in a two-dimensional
  quantum magnet}},}\ }\href {\doibase 10.1103/PhysRevResearch.2.033459}
  {\bibfield  {journal} {\bibinfo  {journal} {Phys. Rev. Res.}\ }\textbf
  {\bibinfo {volume} {2}},\ \bibinfo {pages} {033459} (\bibinfo {year}
  {2020})}\BibitemShut {NoStop}%
\bibitem [{\citenamefont {Ma}\ \emph {et~al.}(2019)\citenamefont {Ma},
  \citenamefont {You},\ and\ \citenamefont {Meng}}]{Ma2019prl}%
  \BibitemOpen
  \bibfield  {author} {\bibinfo {author} {\bibfnamefont {N.}~\bibnamefont
  {Ma}}, \bibinfo {author} {\bibfnamefont {Y.-Z.}\ \bibnamefont {You}}, \ and\
  \bibinfo {author} {\bibfnamefont {Z.~Y.}\ \bibnamefont {Meng}},\ }\bibfield
  {title} {\enquote {\bibinfo {title} {Role of noether's theorem at the
  deconfined quantum critical point},}\ }\href {\doibase
  10.1103/PhysRevLett.122.175701} {\bibfield  {journal} {\bibinfo  {journal}
  {Phys. Rev. Lett.}\ }\textbf {\bibinfo {volume} {122}},\ \bibinfo {pages}
  {175701} (\bibinfo {year} {2019})}\BibitemShut {NoStop}%
\bibitem [{\citenamefont {Amit}\ and\ \citenamefont
  {Peliti}(1982)}]{Amit1982annph}%
  \BibitemOpen
  \bibfield  {author} {\bibinfo {author} {\bibfnamefont {D.~J.}\ \bibnamefont
  {Amit}}\ and\ \bibinfo {author} {\bibfnamefont {L.}~\bibnamefont {Peliti}},\
  }\bibfield  {title} {\enquote {\bibinfo {title} {{On dangerous irrelevant
  operators}},}\ }\href {\doibase https://doi.org/10.1016/0003-4916(82)90159-2}
  {\bibfield  {journal} {\bibinfo  {journal} {Annals of Physics}\ }\textbf
  {\bibinfo {volume} {140}},\ \bibinfo {pages} {207--231} (\bibinfo {year}
  {1982})}\BibitemShut {NoStop}%
\bibitem [{\citenamefont {Nelson}(1976)}]{Nelson1976prb}%
  \BibitemOpen
  \bibfield  {author} {\bibinfo {author} {\bibfnamefont {D.~R.}\ \bibnamefont
  {Nelson}},\ }\bibfield  {title} {\enquote {\bibinfo {title}
  {{Coexistence-curve singularities in isotropic ferromagnets}},}\ }\href
  {\doibase 10.1103/PhysRevB.13.2222} {\bibfield  {journal} {\bibinfo
  {journal} {Phys. Rev. B}\ }\textbf {\bibinfo {volume} {13}},\ \bibinfo
  {pages} {2222--2230} (\bibinfo {year} {1976})}\BibitemShut {NoStop}%
\bibitem [{\citenamefont {Miyashita}(1997)}]{Miyashita1997jpsj}%
  \BibitemOpen
  \bibfield  {author} {\bibinfo {author} {\bibfnamefont {S.}~\bibnamefont
  {Miyashita}},\ }\bibfield  {title} {\enquote {\bibinfo {title} {{Nature of
  the Ordered Phase and the Critical Properties of the Three Dimensional
  Six-State Clock Model}},}\ }\href {\doibase 10.1143/JPSJ.66.3411} {\bibfield
  {journal} {\bibinfo  {journal} {Journal of the Physical Society of Japan}\
  }\textbf {\bibinfo {volume} {66}},\ \bibinfo {pages} {3411--3420} (\bibinfo
  {year} {1997})}\BibitemShut {NoStop}%
\bibitem [{\citenamefont {Shao}\ \emph {et~al.}(2016)\citenamefont {Shao},
  \citenamefont {Guo},\ and\ \citenamefont {Sandvik}}]{Shao2016Sci}%
  \BibitemOpen
  \bibfield  {author} {\bibinfo {author} {\bibfnamefont {H.}~\bibnamefont
  {Shao}}, \bibinfo {author} {\bibfnamefont {W.}~\bibnamefont {Guo}}, \ and\
  \bibinfo {author} {\bibfnamefont {A.~W.}\ \bibnamefont {Sandvik}},\
  }\bibfield  {title} {\enquote {\bibinfo {title} {Quantum criticality with two
  length scales},}\ }\href {\doibase 10.1126/science.aad5007} {\bibfield
  {journal} {\bibinfo  {journal} {Science}\ }\textbf {\bibinfo {volume}
  {352}},\ \bibinfo {pages} {213--216} (\bibinfo {year} {2016})}\BibitemShut
  {NoStop}%
\bibitem [{\citenamefont {Hohenberg}\ and\ \citenamefont
  {Halperin}(1977)}]{Hohenberg1977rmp}%
  \BibitemOpen
  \bibfield  {author} {\bibinfo {author} {\bibfnamefont {P.~C.}\ \bibnamefont
  {Hohenberg}}\ and\ \bibinfo {author} {\bibfnamefont {B.~I.}\ \bibnamefont
  {Halperin}},\ }\bibfield  {title} {\enquote {\bibinfo {title} {{Theory of
  dynamic critical phenomena}},}\ }\href {\doibase 10.1103/RevModPhys.49.435}
  {\bibfield  {journal} {\bibinfo  {journal} {Rev. Mod. Phys.}\ }\textbf
  {\bibinfo {volume} {49}},\ \bibinfo {pages} {435--479} (\bibinfo {year}
  {1977})}\BibitemShut {NoStop}%
\bibitem [{\citenamefont {Folk}\ and\ \citenamefont
  {Moser}(2006)}]{Folk2006jpa}%
  \BibitemOpen
  \bibfield  {author} {\bibinfo {author} {\bibfnamefont {R.}~\bibnamefont
  {Folk}}\ and\ \bibinfo {author} {\bibfnamefont {G.}~\bibnamefont {Moser}},\
  }\bibfield  {title} {\enquote {\bibinfo {title} {{Critical dynamics: a
  field-theoretical approach}},}\ }\href {\doibase 10.1088/0305-4470/39/24/R01}
  {\bibfield  {journal} {\bibinfo  {journal} {Journal of Physics A:
  Mathematical and General}\ }\textbf {\bibinfo {volume} {39}},\ \bibinfo
  {pages} {R207} (\bibinfo {year} {2006})}\BibitemShut {NoStop}%
\bibitem [{\citenamefont {T\"auber}(2014)}]{Tauber2014book}%
  \BibitemOpen
  \bibfield  {author} {\bibinfo {author} {\bibfnamefont {U.~C.}\ \bibnamefont
  {T\"auber}},\ }\href {\doibase 10.1017/CBO9781139046213} {\emph {\bibinfo
  {title} {{Critical Dynamics: A Field Theory Approach to Equilibrium and
  Non-Equilibrium Scaling Behavior}}}}\ (\bibinfo  {publisher} {Cambridge
  University Press},\ \bibinfo {year} {2014})\BibitemShut {NoStop}%
\bibitem [{\citenamefont {Chae}\ \emph {et~al.}(2012)\citenamefont {Chae},
  \citenamefont {Lee}, \citenamefont {Horibe}, \citenamefont {Tanimura},
  \citenamefont {Mori}, \citenamefont {Gao}, \citenamefont {Carr},\ and\
  \citenamefont {Cheong}}]{Chae2012prl}%
  \BibitemOpen
  \bibfield  {author} {\bibinfo {author} {\bibfnamefont {S.~C.}\ \bibnamefont
  {Chae}}, \bibinfo {author} {\bibfnamefont {N.}~\bibnamefont {Lee}}, \bibinfo
  {author} {\bibfnamefont {Y.}~\bibnamefont {Horibe}}, \bibinfo {author}
  {\bibfnamefont {M.}~\bibnamefont {Tanimura}}, \bibinfo {author}
  {\bibfnamefont {S.}~\bibnamefont {Mori}}, \bibinfo {author} {\bibfnamefont
  {B.}~\bibnamefont {Gao}}, \bibinfo {author} {\bibfnamefont {S.}~\bibnamefont
  {Carr}}, \ and\ \bibinfo {author} {\bibfnamefont {S.-W.}\ \bibnamefont
  {Cheong}},\ }\bibfield  {title} {\enquote {\bibinfo {title} {{Direct
  Observation of the Proliferation of Ferroelectric Loop Domains and
  Vortex-Antivortex Pairs}},}\ }\href {\doibase 10.1103/PhysRevLett.108.167603}
  {\bibfield  {journal} {\bibinfo  {journal} {Phys. Rev. Lett.}\ }\textbf
  {\bibinfo {volume} {108}},\ \bibinfo {pages} {167603} (\bibinfo {year}
  {2012})}\BibitemShut {NoStop}%
\bibitem [{\citenamefont {Skj\ae{}rv\o{}}\ \emph {et~al.}(2019)\citenamefont
  {Skj\ae{}rv\o{}}, \citenamefont {Meier}, \citenamefont {Feygenson},
  \citenamefont {Spaldin}, \citenamefont {Billinge}, \citenamefont {Bozin},\
  and\ \citenamefont {Selbach}}]{Skjaervo2019prx}%
  \BibitemOpen
  \bibfield  {author} {\bibinfo {author} {\bibfnamefont {S.~H.}\ \bibnamefont
  {Skj\ae{}rv\o{}}}, \bibinfo {author} {\bibfnamefont {Q.~N.}\ \bibnamefont
  {Meier}}, \bibinfo {author} {\bibfnamefont {M.}~\bibnamefont {Feygenson}},
  \bibinfo {author} {\bibfnamefont {N.~A.}\ \bibnamefont {Spaldin}}, \bibinfo
  {author} {\bibfnamefont {S.~J.~L.}\ \bibnamefont {Billinge}}, \bibinfo
  {author} {\bibfnamefont {E.~S.}\ \bibnamefont {Bozin}}, \ and\ \bibinfo
  {author} {\bibfnamefont {S.~M.}\ \bibnamefont {Selbach}},\ }\bibfield
  {title} {\enquote {\bibinfo {title} {{Unconventional Continuous Structural
  Disorder at the Order-Disorder Phase Transition in the Hexagonal
  Manganites}},}\ }\href {\doibase 10.1103/PhysRevX.9.031001} {\bibfield
  {journal} {\bibinfo  {journal} {Phys. Rev. X}\ }\textbf {\bibinfo {volume}
  {9}},\ \bibinfo {pages} {031001} (\bibinfo {year} {2019})}\BibitemShut
  {NoStop}%
\bibitem [{\citenamefont {Griffin}\ \emph {et~al.}(2012)\citenamefont
  {Griffin}, \citenamefont {Lilienblum}, \citenamefont {Delaney}, \citenamefont
  {Kumagai}, \citenamefont {Fiebig},\ and\ \citenamefont
  {Spaldin}}]{Griffin2012prx}%
  \BibitemOpen
  \bibfield  {author} {\bibinfo {author} {\bibfnamefont {S.~M.}\ \bibnamefont
  {Griffin}}, \bibinfo {author} {\bibfnamefont {M.}~\bibnamefont {Lilienblum}},
  \bibinfo {author} {\bibfnamefont {K.~T.}\ \bibnamefont {Delaney}}, \bibinfo
  {author} {\bibfnamefont {Y.}~\bibnamefont {Kumagai}}, \bibinfo {author}
  {\bibfnamefont {M.}~\bibnamefont {Fiebig}}, \ and\ \bibinfo {author}
  {\bibfnamefont {N.~A.}\ \bibnamefont {Spaldin}},\ }\bibfield  {title}
  {\enquote {\bibinfo {title} {{Scaling Behavior and Beyond Equilibrium in the
  Hexagonal Manganites}},}\ }\href {\doibase 10.1103/PhysRevX.2.041022}
  {\bibfield  {journal} {\bibinfo  {journal} {Phys. Rev. X}\ }\textbf {\bibinfo
  {volume} {2}},\ \bibinfo {pages} {041022} (\bibinfo {year}
  {2012})}\BibitemShut {NoStop}%
\bibitem [{\citenamefont {Lin}\ \emph {et~al.}(2014)\citenamefont {Lin},
  \citenamefont {Wang}, \citenamefont {Kamiya}, \citenamefont {Chern},
  \citenamefont {Fan}, \citenamefont {Fan}, \citenamefont {Casas},
  \citenamefont {Liu}, \citenamefont {Kiryukhin}, \citenamefont {Zurek},
  \citenamefont {Batista},\ and\ \citenamefont {Cheong}}]{Lin2014natphy}%
  \BibitemOpen
  \bibfield  {author} {\bibinfo {author} {\bibfnamefont {S.-Z.}\ \bibnamefont
  {Lin}}, \bibinfo {author} {\bibfnamefont {X.}~\bibnamefont {Wang}}, \bibinfo
  {author} {\bibfnamefont {Y.}~\bibnamefont {Kamiya}}, \bibinfo {author}
  {\bibfnamefont {G.-W.}\ \bibnamefont {Chern}}, \bibinfo {author}
  {\bibfnamefont {F.}~\bibnamefont {Fan}}, \bibinfo {author} {\bibfnamefont
  {D.}~\bibnamefont {Fan}}, \bibinfo {author} {\bibfnamefont {B.}~\bibnamefont
  {Casas}}, \bibinfo {author} {\bibfnamefont {Y.}~\bibnamefont {Liu}}, \bibinfo
  {author} {\bibfnamefont {V.}~\bibnamefont {Kiryukhin}}, \bibinfo {author}
  {\bibfnamefont {W.~H.}\ \bibnamefont {Zurek}}, \bibinfo {author}
  {\bibfnamefont {C.~D.}\ \bibnamefont {Batista}}, \ and\ \bibinfo {author}
  {\bibfnamefont {S.-W.}\ \bibnamefont {Cheong}},\ }\bibfield  {title}
  {\enquote {\bibinfo {title} {{Topological defects as relics of emergent
  continuous symmetry and Higgs condensation of disorder in
  ferroelectrics}},}\ }\href {\doibase 10.1038/nphys3142} {\bibfield  {journal}
  {\bibinfo  {journal} {Nature Physics}\ }\textbf {\bibinfo {volume} {10}},\
  \bibinfo {pages} {970--977} (\bibinfo {year} {2014})}\BibitemShut {NoStop}%
\bibitem [{\citenamefont {Meier}\ \emph {et~al.}(2017)\citenamefont {Meier},
  \citenamefont {Lilienblum}, \citenamefont {Griffin}, \citenamefont {Conder},
  \citenamefont {Pomjakushina}, \citenamefont {Yan}, \citenamefont {Bourret},
  \citenamefont {Meier}, \citenamefont {Lichtenberg}, \citenamefont {Salje},
  \citenamefont {Spaldin}, \citenamefont {Fiebig},\ and\ \citenamefont
  {Cano}}]{Meier2017prx}%
  \BibitemOpen
  \bibfield  {author} {\bibinfo {author} {\bibfnamefont {Q.~N.}\ \bibnamefont
  {Meier}}, \bibinfo {author} {\bibfnamefont {M.}~\bibnamefont {Lilienblum}},
  \bibinfo {author} {\bibfnamefont {S.~M.}\ \bibnamefont {Griffin}}, \bibinfo
  {author} {\bibfnamefont {K.}~\bibnamefont {Conder}}, \bibinfo {author}
  {\bibfnamefont {E.}~\bibnamefont {Pomjakushina}}, \bibinfo {author}
  {\bibfnamefont {Z.}~\bibnamefont {Yan}}, \bibinfo {author} {\bibfnamefont
  {E.}~\bibnamefont {Bourret}}, \bibinfo {author} {\bibfnamefont
  {D.}~\bibnamefont {Meier}}, \bibinfo {author} {\bibfnamefont
  {F.}~\bibnamefont {Lichtenberg}}, \bibinfo {author} {\bibfnamefont
  {E.~K.~H.}\ \bibnamefont {Salje}}, \bibinfo {author} {\bibfnamefont {N.~A.}\
  \bibnamefont {Spaldin}}, \bibinfo {author} {\bibfnamefont {M.}~\bibnamefont
  {Fiebig}}, \ and\ \bibinfo {author} {\bibfnamefont {A.}~\bibnamefont
  {Cano}},\ }\bibfield  {title} {\enquote {\bibinfo {title} {{Global Formation
  of Topological Defects in the Multiferroic Hexagonal Manganites}},}\ }\href
  {\doibase 10.1103/PhysRevX.7.041014} {\bibfield  {journal} {\bibinfo
  {journal} {Phys. Rev. X}\ }\textbf {\bibinfo {volume} {7}},\ \bibinfo {pages}
  {041014} (\bibinfo {year} {2017})}\BibitemShut {NoStop}%
\bibitem [{\citenamefont {Meier}\ \emph {et~al.}(2020)\citenamefont {Meier},
  \citenamefont {Stucky}, \citenamefont {Teyssier}, \citenamefont {Griffin},
  \citenamefont {van~der Marel},\ and\ \citenamefont {Spaldin}}]{Meier2020prb}%
  \BibitemOpen
  \bibfield  {author} {\bibinfo {author} {\bibfnamefont {Q.~N.}\ \bibnamefont
  {Meier}}, \bibinfo {author} {\bibfnamefont {A.}~\bibnamefont {Stucky}},
  \bibinfo {author} {\bibfnamefont {J.}~\bibnamefont {Teyssier}}, \bibinfo
  {author} {\bibfnamefont {S.~M.}\ \bibnamefont {Griffin}}, \bibinfo {author}
  {\bibfnamefont {D.}~\bibnamefont {van~der Marel}}, \ and\ \bibinfo {author}
  {\bibfnamefont {N.~A.}\ \bibnamefont {Spaldin}},\ }\bibfield  {title}
  {\enquote {\bibinfo {title} {Manifestation of structural higgs and goldstone
  modes in the hexagonal manganites},}\ }\href {\doibase
  10.1103/PhysRevB.102.014102} {\bibfield  {journal} {\bibinfo  {journal}
  {Phys. Rev. B}\ }\textbf {\bibinfo {volume} {102}},\ \bibinfo {pages}
  {014102} (\bibinfo {year} {2020})}\BibitemShut {NoStop}%
\bibitem [{\citenamefont {Zhang}\ \emph {et~al.}(2021)\citenamefont {Zhang},
  \citenamefont {Ye},\ and\ \citenamefont {Li}}]{Zhang2021prb}%
  \BibitemOpen
  \bibfield  {author} {\bibinfo {author} {\bibfnamefont {X.}~\bibnamefont
  {Zhang}}, \bibinfo {author} {\bibfnamefont {Q.-J.}\ \bibnamefont {Ye}}, \
  and\ \bibinfo {author} {\bibfnamefont {X.-Z.}\ \bibnamefont {Li}},\
  }\bibfield  {title} {\enquote {\bibinfo {title} {{Structural phase transition
  and Goldstone-like mode in hexagonal ${\mathrm{BaMnO}}_{3}$}},}\ }\href
  {\doibase 10.1103/PhysRevB.103.024101} {\bibfield  {journal} {\bibinfo
  {journal} {Phys. Rev. B}\ }\textbf {\bibinfo {volume} {103}},\ \bibinfo
  {pages} {024101} (\bibinfo {year} {2021})}\BibitemShut {NoStop}%
\bibitem [{\citenamefont {Sandvik}\ \emph {et~al.}(2023)\citenamefont
  {Sandvik}, \citenamefont {M\"u{}ller}, \citenamefont {\AA{}nes},
  \citenamefont {Zahn}, \citenamefont {He}, \citenamefont {Fiebig},
  \citenamefont {Lottermoser}, \citenamefont {Rojac}, \citenamefont {Meier},\
  and\ \citenamefont {Schulthei\ss}}]{OWSandvik2023nl}%
  \BibitemOpen
  \bibfield  {author} {\bibinfo {author} {\bibfnamefont {O.~W.}\ \bibnamefont
  {Sandvik}}, \bibinfo {author} {\bibfnamefont {A.~M.}\ \bibnamefont
  {M\"u{}ller}}, \bibinfo {author} {\bibfnamefont {H.~W.}\ \bibnamefont
  {\AA{}nes}}, \bibinfo {author} {\bibfnamefont {M.}~\bibnamefont {Zahn}},
  \bibinfo {author} {\bibfnamefont {J.}~\bibnamefont {He}}, \bibinfo {author}
  {\bibfnamefont {M.}~\bibnamefont {Fiebig}}, \bibinfo {author} {\bibfnamefont
  {T.}~\bibnamefont {Lottermoser}}, \bibinfo {author} {\bibfnamefont
  {T.}~\bibnamefont {Rojac}}, \bibinfo {author} {\bibfnamefont
  {D.}~\bibnamefont {Meier}}, \ and\ \bibinfo {author} {\bibfnamefont
  {J.}~\bibnamefont {Schulthei\ss}},\ }\bibfield  {title} {\enquote {\bibinfo
  {title} {{Pressure Control of Nonferroelastic Ferroelectric Domains in
  ErMnO$_3$}},}\ }\href {\doibase 10.1021/acs.nanolett.3c01638} {\bibfield
  {journal} {\bibinfo  {journal} {Nano Letters}\ }\textbf {\bibinfo {volume}
  {23}},\ \bibinfo {pages} {6994--7000} (\bibinfo {year} {2023})}\BibitemShut
  {NoStop}%
\bibitem [{\citenamefont {Kang}\ \emph {et~al.}(2023)\citenamefont {Kang},
  \citenamefont {Gao}, \citenamefont {Guo}, \citenamefont {Zhu}, \citenamefont
  {Huang}, \citenamefont {Hong}, \citenamefont {Cheong},\ and\ \citenamefont
  {Wang}}]{Kang2023jap}%
  \BibitemOpen
  \bibfield  {author} {\bibinfo {author} {\bibfnamefont {J.}~\bibnamefont
  {Kang}}, \bibinfo {author} {\bibfnamefont {Z.}~\bibnamefont {Gao}}, \bibinfo
  {author} {\bibfnamefont {C.}~\bibnamefont {Guo}}, \bibinfo {author}
  {\bibfnamefont {W.}~\bibnamefont {Zhu}}, \bibinfo {author} {\bibfnamefont
  {H.}~\bibnamefont {Huang}}, \bibinfo {author} {\bibfnamefont
  {J.}~\bibnamefont {Hong}}, \bibinfo {author} {\bibfnamefont {S.-W.}\
  \bibnamefont {Cheong}}, \ and\ \bibinfo {author} {\bibfnamefont
  {X.}~\bibnamefont {Wang}},\ }\bibfield  {title} {\enquote {\bibinfo {title}
  {{A snapshot of domain evolution between topological vortex and stripe in
  ferroelectric hexagonal ErMnO3}},}\ }\href {\doibase 10.1063/5.0138096}
  {\bibfield  {journal} {\bibinfo  {journal} {Journal of Applied Physics}\
  }\textbf {\bibinfo {volume} {133}},\ \bibinfo {pages} {124102} (\bibinfo
  {year} {2023})}\BibitemShut {NoStop}%
\bibitem [{\citenamefont {Baghizadeh}\ \emph {et~al.}(2019)\citenamefont
  {Baghizadeh}, \citenamefont {Mirzadeh~Vaghefi}, \citenamefont {Alikin},
  \citenamefont {Amaral}, \citenamefont {Amaral},\ and\ \citenamefont
  {Vieira}}]{Baghizadeh2019jpcc}%
  \BibitemOpen
  \bibfield  {author} {\bibinfo {author} {\bibfnamefont {A.}~\bibnamefont
  {Baghizadeh}}, \bibinfo {author} {\bibfnamefont {P.}~\bibnamefont
  {Mirzadeh~Vaghefi}}, \bibinfo {author} {\bibfnamefont {D.~O.}\ \bibnamefont
  {Alikin}}, \bibinfo {author} {\bibfnamefont {J.~S.}\ \bibnamefont {Amaral}},
  \bibinfo {author} {\bibfnamefont {V.~S.}\ \bibnamefont {Amaral}}, \ and\
  \bibinfo {author} {\bibfnamefont {J.~M.}\ \bibnamefont {Vieira}},\ }\bibfield
   {title} {\enquote {\bibinfo {title} {{Link of Weak Ferromagnetism to
  Emergence of Topological Vortices in Bulk Ceramics of
  h-${\mathrm{LuMn}}_{x}{\mathrm{O}}_{3}$ Manganite}},}\ }\href {\doibase
  10.1021/acs.jpcc.8b11253} {\bibfield  {journal} {\bibinfo  {journal} {The
  Journal of Physical Chemistry C}\ }\textbf {\bibinfo {volume} {123}},\
  \bibinfo {pages} {6158--6166} (\bibinfo {year} {2019})}\BibitemShut {NoStop}%
\bibitem [{\citenamefont {Juraschek}\ \emph {et~al.}(2020)\citenamefont
  {Juraschek}, \citenamefont {Meier},\ and\ \citenamefont
  {Narang}}]{Juraschek2020prl}%
  \BibitemOpen
  \bibfield  {author} {\bibinfo {author} {\bibfnamefont {D.~M.}\ \bibnamefont
  {Juraschek}}, \bibinfo {author} {\bibfnamefont {Q.~N.}\ \bibnamefont
  {Meier}}, \ and\ \bibinfo {author} {\bibfnamefont {P.}~\bibnamefont
  {Narang}},\ }\bibfield  {title} {\enquote {\bibinfo {title} {Parametric
  excitation of an optically silent goldstone-like phonon mode},}\ }\href
  {\doibase 10.1103/PhysRevLett.124.117401} {\bibfield  {journal} {\bibinfo
  {journal} {Phys. Rev. Lett.}\ }\textbf {\bibinfo {volume} {124}},\ \bibinfo
  {pages} {117401} (\bibinfo {year} {2020})}\BibitemShut {NoStop}%
\bibitem [{\citenamefont {Adzhemyan}\ \emph {et~al.}(2022)\citenamefont
  {Adzhemyan}, \citenamefont {Evdokimov}, \citenamefont {Hnatič},
  \citenamefont {Ivanova}, \citenamefont {Kompaniets}, \citenamefont {Kudlis},\
  and\ \citenamefont {Zakharov}}]{Adzhemyan2022pa}%
  \BibitemOpen
  \bibfield  {author} {\bibinfo {author} {\bibfnamefont {L.}~\bibnamefont
  {Adzhemyan}}, \bibinfo {author} {\bibfnamefont {D.}~\bibnamefont
  {Evdokimov}}, \bibinfo {author} {\bibfnamefont {M.}~\bibnamefont {Hnatič}},
  \bibinfo {author} {\bibfnamefont {E.}~\bibnamefont {Ivanova}}, \bibinfo
  {author} {\bibfnamefont {M.}~\bibnamefont {Kompaniets}}, \bibinfo {author}
  {\bibfnamefont {A.}~\bibnamefont {Kudlis}}, \ and\ \bibinfo {author}
  {\bibfnamefont {D.}~\bibnamefont {Zakharov}},\ }\bibfield  {title} {\enquote
  {\bibinfo {title} {Model a of critical dynamics: 5-loop $\varepsilon$
  expansion study},}\ }\href {\doibase
  https://doi.org/10.1016/j.physa.2022.127530} {\bibfield  {journal} {\bibinfo
  {journal} {Physica A: Statistical Mechanics and its Applications}\ }\textbf
  {\bibinfo {volume} {600}},\ \bibinfo {pages} {127530} (\bibinfo {year}
  {2022})}\BibitemShut {NoStop}%
\bibitem [{\citenamefont {Xiang}\ \emph {et~al.}(2024)\citenamefont {Xiang},
  \citenamefont {Zhang}, \citenamefont {Gao}, \citenamefont {Schmidt},
  \citenamefont {Schmalzl}, \citenamefont {Wang}, \citenamefont {Li},
  \citenamefont {Xi}, \citenamefont {Liu}, \citenamefont {Jin}, \citenamefont
  {Li}, \citenamefont {Shen}, \citenamefont {Chen}, \citenamefont {Qi},
  \citenamefont {Wan}, \citenamefont {Jin}, \citenamefont {Li}, \citenamefont
  {Sun},\ and\ \citenamefont {Su}}]{Xiang2024nat}%
  \BibitemOpen
  \bibfield  {author} {\bibinfo {author} {\bibfnamefont {J.}~\bibnamefont
  {Xiang}}, \bibinfo {author} {\bibfnamefont {C.}~\bibnamefont {Zhang}},
  \bibinfo {author} {\bibfnamefont {Y.}~\bibnamefont {Gao}}, \bibinfo {author}
  {\bibfnamefont {W.}~\bibnamefont {Schmidt}}, \bibinfo {author} {\bibfnamefont
  {K.}~\bibnamefont {Schmalzl}}, \bibinfo {author} {\bibfnamefont {C.-W.}\
  \bibnamefont {Wang}}, \bibinfo {author} {\bibfnamefont {B.}~\bibnamefont
  {Li}}, \bibinfo {author} {\bibfnamefont {N.}~\bibnamefont {Xi}}, \bibinfo
  {author} {\bibfnamefont {X.-Y.}\ \bibnamefont {Liu}}, \bibinfo {author}
  {\bibfnamefont {H.}~\bibnamefont {Jin}}, \bibinfo {author} {\bibfnamefont
  {G.}~\bibnamefont {Li}}, \bibinfo {author} {\bibfnamefont {J.}~\bibnamefont
  {Shen}}, \bibinfo {author} {\bibfnamefont {Z.}~\bibnamefont {Chen}}, \bibinfo
  {author} {\bibfnamefont {Y.}~\bibnamefont {Qi}}, \bibinfo {author}
  {\bibfnamefont {Y.}~\bibnamefont {Wan}}, \bibinfo {author} {\bibfnamefont
  {W.}~\bibnamefont {Jin}}, \bibinfo {author} {\bibfnamefont {W.}~\bibnamefont
  {Li}}, \bibinfo {author} {\bibfnamefont {P.}~\bibnamefont {Sun}}, \ and\
  \bibinfo {author} {\bibfnamefont {G.}~\bibnamefont {Su}},\ }\bibfield
  {title} {\enquote {\bibinfo {title} {{Giant magnetocaloric effect in spin
  supersolid candidate ${\mathrm{Na}_{2}}{\mathrm{BaCo(PO}}_{4})_{2}$}},}\
  }\href {\doibase 10.1038/s41586-023-06885-w} {\bibfield  {journal} {\bibinfo
  {journal} {Nature}\ }\textbf {\bibinfo {volume} {625}},\ \bibinfo {pages}
  {270--275} (\bibinfo {year} {2024})}\BibitemShut {NoStop}%
\bibitem [{\citenamefont {Chi}\ \emph {et~al.}(2024)\citenamefont {Chi},
  \citenamefont {Hu}, \citenamefont {Liao},\ and\ \citenamefont
  {Xiang}}]{Chi2024}%
  \BibitemOpen
  \bibfield  {author} {\bibinfo {author} {\bibfnamefont {R.}~\bibnamefont
  {Chi}}, \bibinfo {author} {\bibfnamefont {J.}~\bibnamefont {Hu}}, \bibinfo
  {author} {\bibfnamefont {H.-J.}\ \bibnamefont {Liao}}, \ and\ \bibinfo
  {author} {\bibfnamefont {T.}~\bibnamefont {Xiang}},\ }\bibfield  {title}
  {\enquote {\bibinfo {title} {{Dynamical Spectra of Spin Supersolid States in
  Triangular Antiferromagnets}},}\ }\href {https://arxiv.org/abs/2404.14163}
  {\bibfield  {journal} {\bibinfo  {journal} {arXiv:2024.14163}\ } (\bibinfo
  {year} {2024})}\BibitemShut {NoStop}%
\bibitem [{\citenamefont {Gao}\ \emph {et~al.}()\citenamefont {Gao},
  \citenamefont {Zhang}, \citenamefont {Xiang}, \citenamefont {Yu},
  \citenamefont {Lu}, \citenamefont {Sun}, \citenamefont {Jin}, \citenamefont
  {Su},\ and\ \citenamefont {Li}}]{Gao2024}%
  \BibitemOpen
  \bibfield  {author} {\bibinfo {author} {\bibfnamefont {Y.}~\bibnamefont
  {Gao}}, \bibinfo {author} {\bibfnamefont {C.}~\bibnamefont {Zhang}}, \bibinfo
  {author} {\bibfnamefont {J.}~\bibnamefont {Xiang}}, \bibinfo {author}
  {\bibfnamefont {D.}~\bibnamefont {Yu}}, \bibinfo {author} {\bibfnamefont
  {X.}~\bibnamefont {Lu}}, \bibinfo {author} {\bibfnamefont {P.}~\bibnamefont
  {Sun}}, \bibinfo {author} {\bibfnamefont {W.}~\bibnamefont {Jin}}, \bibinfo
  {author} {\bibfnamefont {G.}~\bibnamefont {Su}}, \ and\ \bibinfo {author}
  {\bibfnamefont {W.}~\bibnamefont {Li}},\ }\bibfield  {title} {\enquote
  {\bibinfo {title} {Spin supersolid phase and double magnon-roton excitations
  in a cobalt-based triangular lattice},}\ }\href@noop {} {\ }\BibitemShut
  {NoStop}%
\bibitem [{\citenamefont {Campostrini}\ \emph {et~al.}(2006)\citenamefont
  {Campostrini}, \citenamefont {Hasenbusch}, \citenamefont {Pelissetto},\ and\
  \citenamefont {Vicari}}]{Campostrini2006prb}%
  \BibitemOpen
  \bibfield  {author} {\bibinfo {author} {\bibfnamefont {M.}~\bibnamefont
  {Campostrini}}, \bibinfo {author} {\bibfnamefont {M.}~\bibnamefont
  {Hasenbusch}}, \bibinfo {author} {\bibfnamefont {A.}~\bibnamefont
  {Pelissetto}}, \ and\ \bibinfo {author} {\bibfnamefont {E.}~\bibnamefont
  {Vicari}},\ }\bibfield  {title} {\enquote {\bibinfo {title} {{Theoretical
  estimates of the critical exponents of the superfluid transition in
  $^{4}\mathrm{He}$ by lattice methods}},}\ }\href {\doibase
  10.1103/PhysRevB.74.144506} {\bibfield  {journal} {\bibinfo  {journal} {Phys.
  Rev. B}\ }\textbf {\bibinfo {volume} {74}},\ \bibinfo {pages} {144506}
  (\bibinfo {year} {2006})}\BibitemShut {NoStop}%
\bibitem [{\citenamefont {Campostrini}\ \emph {et~al.}(2001)\citenamefont
  {Campostrini}, \citenamefont {Hasenbusch}, \citenamefont {Pelissetto},
  \citenamefont {Rossi},\ and\ \citenamefont {Vicari}}]{Campostrini2001prb}%
  \BibitemOpen
  \bibfield  {author} {\bibinfo {author} {\bibfnamefont {M.}~\bibnamefont
  {Campostrini}}, \bibinfo {author} {\bibfnamefont {M.}~\bibnamefont
  {Hasenbusch}}, \bibinfo {author} {\bibfnamefont {A.}~\bibnamefont
  {Pelissetto}}, \bibinfo {author} {\bibfnamefont {P.}~\bibnamefont {Rossi}}, \
  and\ \bibinfo {author} {\bibfnamefont {E.}~\bibnamefont {Vicari}},\
  }\bibfield  {title} {\enquote {\bibinfo {title} {{Critical behavior of the
  three-dimensional $\mathrm{XY}$ universality class}},}\ }\href {\doibase
  10.1103/PhysRevB.63.214503} {\bibfield  {journal} {\bibinfo  {journal} {Phys.
  Rev. B}\ }\textbf {\bibinfo {volume} {63}},\ \bibinfo {pages} {214503}
  (\bibinfo {year} {2001})}\BibitemShut {NoStop}%
\bibitem [{\citenamefont {Chester}\ \emph {et~al.}(2020)\citenamefont
  {Chester}, \citenamefont {Landry}, \citenamefont {Liu}, \citenamefont
  {Poland}, \citenamefont {Simmons-Duffin}, \citenamefont {Su},\ and\
  \citenamefont {Vichi}}]{Chester2020jhep}%
  \BibitemOpen
  \bibfield  {author} {\bibinfo {author} {\bibfnamefont {S.~M.}\ \bibnamefont
  {Chester}}, \bibinfo {author} {\bibfnamefont {W.}~\bibnamefont {Landry}},
  \bibinfo {author} {\bibfnamefont {J.}~\bibnamefont {Liu}}, \bibinfo {author}
  {\bibfnamefont {D.}~\bibnamefont {Poland}}, \bibinfo {author} {\bibfnamefont
  {D.}~\bibnamefont {Simmons-Duffin}}, \bibinfo {author} {\bibfnamefont
  {N.}~\bibnamefont {Su}}, \ and\ \bibinfo {author} {\bibfnamefont
  {A.}~\bibnamefont {Vichi}},\ }\bibfield  {title} {\enquote {\bibinfo {title}
  {{Carving out OPE space and precise $O(2)$ model critical exponents}},}\
  }\href {\doibase 10.1007/JHEP06(2020)142} {\bibfield  {journal} {\bibinfo
  {journal} {Journal of High Energy Physics}\ }\textbf {\bibinfo {volume}
  {2020}},\ \bibinfo {pages} {142} (\bibinfo {year} {2020})}\BibitemShut
  {NoStop}%
\bibitem [{\citenamefont {Ueno}\ and\ \citenamefont
  {Mitsubo}(1991)}]{Ueno1991prb}%
  \BibitemOpen
  \bibfield  {author} {\bibinfo {author} {\bibfnamefont {Y.}~\bibnamefont
  {Ueno}}\ and\ \bibinfo {author} {\bibfnamefont {K.}~\bibnamefont {Mitsubo}},\
  }\bibfield  {title} {\enquote {\bibinfo {title} {{Incompletely ordered phase
  in the three-dimensional six-state clock model: Evidence for an absence of
  ordered phases of XY character}},}\ }\href {\doibase
  10.1103/PhysRevB.43.8654} {\bibfield  {journal} {\bibinfo  {journal} {Phys.
  Rev. B}\ }\textbf {\bibinfo {volume} {43}},\ \bibinfo {pages} {8654--8657}
  (\bibinfo {year} {1991})}\BibitemShut {NoStop}%
\bibitem [{\citenamefont {Chubukov}\ \emph {et~al.}(1994)\citenamefont
  {Chubukov}, \citenamefont {Sachdev},\ and\ \citenamefont
  {Ye}}]{Chubukov1994prb}%
  \BibitemOpen
  \bibfield  {author} {\bibinfo {author} {\bibfnamefont {A.~V.}\ \bibnamefont
  {Chubukov}}, \bibinfo {author} {\bibfnamefont {S.}~\bibnamefont {Sachdev}}, \
  and\ \bibinfo {author} {\bibfnamefont {J.}~\bibnamefont {Ye}},\ }\bibfield
  {title} {\enquote {\bibinfo {title} {{Theory of two-dimensional quantum
  Heisenberg antiferromagnets with a nearly critical ground state}},}\ }\href
  {\doibase 10.1103/PhysRevB.49.11919} {\bibfield  {journal} {\bibinfo
  {journal} {Phys. Rev. B}\ }\textbf {\bibinfo {volume} {49}},\ \bibinfo
  {pages} {11919--11961} (\bibinfo {year} {1994})}\BibitemShut {NoStop}%
\bibitem [{\citenamefont {Banerjee}\ \emph {et~al.}(2018)\citenamefont
  {Banerjee}, \citenamefont {Chandrasekharan},\ and\ \citenamefont
  {Orlando}}]{Banerjee2018prl}%
  \BibitemOpen
  \bibfield  {author} {\bibinfo {author} {\bibfnamefont {D.}~\bibnamefont
  {Banerjee}}, \bibinfo {author} {\bibfnamefont {S.}~\bibnamefont
  {Chandrasekharan}}, \ and\ \bibinfo {author} {\bibfnamefont {D.}~\bibnamefont
  {Orlando}},\ }\bibfield  {title} {\enquote {\bibinfo {title} {{Conformal
  Dimensions via Large Charge Expansion}},}\ }\href {\doibase
  10.1103/PhysRevLett.120.061603} {\bibfield  {journal} {\bibinfo  {journal}
  {Phys. Rev. Lett.}\ }\textbf {\bibinfo {volume} {120}},\ \bibinfo {pages}
  {061603} (\bibinfo {year} {2018})}\BibitemShut {NoStop}%
\bibitem [{\citenamefont {Binder}\ and\ \citenamefont
  {Heermann}(2010)}]{binderbook}%
  \BibitemOpen
  \bibfield  {author} {\bibinfo {author} {\bibfnamefont {K.}~\bibnamefont
  {Binder}}\ and\ \bibinfo {author} {\bibfnamefont {D.~W.}\ \bibnamefont
  {Heermann}},\ }\href@noop {} {\emph {\bibinfo {title} {{Monte Carlo
  Simulation in Statistical Physics}}}}\ (\bibinfo  {publisher} {Springer
  Berlin, Heidelberg},\ \bibinfo {year} {2010})\BibitemShut {NoStop}%
\bibitem [{\citenamefont {Fedorenko}\ and\ \citenamefont
  {Trimper}(2006)}]{Fedorenko2006epl}%
  \BibitemOpen
  \bibfield  {author} {\bibinfo {author} {\bibfnamefont {A.~A.}\ \bibnamefont
  {Fedorenko}}\ and\ \bibinfo {author} {\bibfnamefont {S.}~\bibnamefont
  {Trimper}},\ }\bibfield  {title} {\enquote {\bibinfo {title} {Critical aging
  of a ferromagnetic system from a completely ordered state},}\ }\href
  {\doibase 10.1209/epl/i2005-10500-9} {\bibfield  {journal} {\bibinfo
  {journal} {Europhysics Letters}\ }\textbf {\bibinfo {volume} {74}},\ \bibinfo
  {pages} {89} (\bibinfo {year} {2006})}\BibitemShut {NoStop}%
\bibitem [{\citenamefont {Zheng}(1998)}]{Zhengb1998ijmpb}%
  \BibitemOpen
  \bibfield  {author} {\bibinfo {author} {\bibfnamefont {B.}~\bibnamefont
  {Zheng}},\ }\bibfield  {title} {\enquote {\bibinfo {title} {Monte carlo
  simulations of short-time critical dynamics},}\ }\href {\doibase
  10.1142/S021797929800288X} {\bibfield  {journal} {\bibinfo  {journal}
  {International Journal of Modern Physics B}\ }\textbf {\bibinfo {volume}
  {12}},\ \bibinfo {pages} {1419--1484} (\bibinfo {year} {1998})}\BibitemShut
  {NoStop}%
\bibitem [{\citenamefont {Luo}\ \emph {et~al.}(1998)\citenamefont {Luo},
  \citenamefont {Sch\"ulke},\ and\ \citenamefont {Zheng}}]{Luo1998prl}%
  \BibitemOpen
  \bibfield  {author} {\bibinfo {author} {\bibfnamefont {H.~J.}\ \bibnamefont
  {Luo}}, \bibinfo {author} {\bibfnamefont {L.}~\bibnamefont {Sch\"ulke}}, \
  and\ \bibinfo {author} {\bibfnamefont {B.}~\bibnamefont {Zheng}},\ }\bibfield
   {title} {\enquote {\bibinfo {title} {Dynamic approach to the fully
  frustrated $\mathit{XY}$ model},}\ }\href {\doibase
  10.1103/PhysRevLett.81.180} {\bibfield  {journal} {\bibinfo  {journal} {Phys.
  Rev. Lett.}\ }\textbf {\bibinfo {volume} {81}},\ \bibinfo {pages} {180--183}
  (\bibinfo {year} {1998})}\BibitemShut {NoStop}%
\bibitem [{\citenamefont {Stauffer}(1992)}]{Stauffer1992pa}%
  \BibitemOpen
  \bibfield  {author} {\bibinfo {author} {\bibfnamefont {D.}~\bibnamefont
  {Stauffer}},\ }\bibfield  {title} {\enquote {\bibinfo {title} {Kinetics of
  clusters in ising models},}\ }\href
  {https://EconPapers.repec.org/RePEc:eee:phsmap:v:186:y:1992:i:1:p:197-209}
  {\bibfield  {journal} {\bibinfo  {journal} {Physica A: Statistical Mechanics
  and its Applications}\ }\textbf {\bibinfo {volume} {186}},\ \bibinfo {pages}
  {197--209} (\bibinfo {year} {1992})}\BibitemShut {NoStop}%
\bibitem [{\citenamefont {Ito}(1993)}]{Ito1993pa}%
  \BibitemOpen
  \bibfield  {author} {\bibinfo {author} {\bibfnamefont {N.}~\bibnamefont
  {Ito}},\ }\bibfield  {title} {\enquote {\bibinfo {title} {Non-equilibrium
  relaxation and interface energy of the ising model},}\ }\href
  {https://EconPapers.repec.org/RePEc:eee:phsmap:v:196:y:1993:i:4:p:591-614}
  {\bibfield  {journal} {\bibinfo  {journal} {Physica A: Statistical Mechanics
  and its Applications}\ }\textbf {\bibinfo {volume} {196}},\ \bibinfo {pages}
  {591--614} (\bibinfo {year} {1993})}\BibitemShut {NoStop}%
\bibitem [{\citenamefont {Albano}\ \emph {et~al.}(2011)\citenamefont {Albano},
  \citenamefont {Bab}, \citenamefont {Baglietto}, \citenamefont {Borzi},
  \citenamefont {Grigera}, \citenamefont {Loscar}, \citenamefont {Rodriguez},
  \citenamefont {Puzzo},\ and\ \citenamefont {Saracco}}]{Albano2011rpp}%
  \BibitemOpen
  \bibfield  {author} {\bibinfo {author} {\bibfnamefont {E.~V.}\ \bibnamefont
  {Albano}}, \bibinfo {author} {\bibfnamefont {M.~A.}\ \bibnamefont {Bab}},
  \bibinfo {author} {\bibfnamefont {G.}~\bibnamefont {Baglietto}}, \bibinfo
  {author} {\bibfnamefont {R.~A.}\ \bibnamefont {Borzi}}, \bibinfo {author}
  {\bibfnamefont {T.~S.}\ \bibnamefont {Grigera}}, \bibinfo {author}
  {\bibfnamefont {E.~S.}\ \bibnamefont {Loscar}}, \bibinfo {author}
  {\bibfnamefont {D.~E.}\ \bibnamefont {Rodriguez}}, \bibinfo {author}
  {\bibfnamefont {M.~L.~R.}\ \bibnamefont {Puzzo}}, \ and\ \bibinfo {author}
  {\bibfnamefont {G.~P.}\ \bibnamefont {Saracco}},\ }\bibfield  {title}
  {\enquote {\bibinfo {title} {Study of phase transitions from short-time
  non-equilibrium behaviour},}\ }\href {\doibase 10.1088/0034-4885/74/2/026501}
  {\bibfield  {journal} {\bibinfo  {journal} {Reports on Progress in Physics}\
  }\textbf {\bibinfo {volume} {74}},\ \bibinfo {pages} {026501} (\bibinfo
  {year} {2011})}\BibitemShut {NoStop}%
\bibitem [{\citenamefont {Huang}\ \emph {et~al.}(2021)\citenamefont {Huang},
  \citenamefont {Li}, \citenamefont {Xue}, \citenamefont {Kim}, \citenamefont
  {Zhang}, \citenamefont {Chu}, \citenamefont {Chen},\ and\ \citenamefont
  {Cheong}}]{Huang2021prr}%
  \BibitemOpen
  \bibfield  {author} {\bibinfo {author} {\bibfnamefont {F.-T.}\ \bibnamefont
  {Huang}}, \bibinfo {author} {\bibfnamefont {Y.}~\bibnamefont {Li}}, \bibinfo
  {author} {\bibfnamefont {F.}~\bibnamefont {Xue}}, \bibinfo {author}
  {\bibfnamefont {J.-W.}\ \bibnamefont {Kim}}, \bibinfo {author} {\bibfnamefont
  {L.}~\bibnamefont {Zhang}}, \bibinfo {author} {\bibfnamefont {M.-W.}\
  \bibnamefont {Chu}}, \bibinfo {author} {\bibfnamefont {L.-Q.}\ \bibnamefont
  {Chen}}, \ and\ \bibinfo {author} {\bibfnamefont {S.-W.}\ \bibnamefont
  {Cheong}},\ }\bibfield  {title} {\enquote {\bibinfo {title} {{Evolution of
  topological defects at two sequential phase transitions of
  ${\mathrm{Nd}}_{2}\mathrm{Sr}{\mathrm{Fe}}_{2}{\mathrm{O}}_{7}$}},}\ }\href
  {\doibase 10.1103/PhysRevResearch.3.023216} {\bibfield  {journal} {\bibinfo
  {journal} {Phys. Rev. Res.}\ }\textbf {\bibinfo {volume} {3}},\ \bibinfo
  {pages} {023216} (\bibinfo {year} {2021})}\BibitemShut {NoStop}%
\bibitem [{\citenamefont {Xu}\ \emph {et~al.}(2022)\citenamefont {Xu},
  \citenamefont {Huang}, \citenamefont {Admasu}, \citenamefont {Kim},
  \citenamefont {Wang}, \citenamefont {Feng}, \citenamefont {Cao},\ and\
  \citenamefont {Cheong}}]{Xu2022cm}%
  \BibitemOpen
  \bibfield  {author} {\bibinfo {author} {\bibfnamefont {X.}~\bibnamefont
  {Xu}}, \bibinfo {author} {\bibfnamefont {F.-T.}\ \bibnamefont {Huang}},
  \bibinfo {author} {\bibfnamefont {A.~S.}\ \bibnamefont {Admasu}}, \bibinfo
  {author} {\bibfnamefont {J.}~\bibnamefont {Kim}}, \bibinfo {author}
  {\bibfnamefont {K.}~\bibnamefont {Wang}}, \bibinfo {author} {\bibfnamefont
  {E.}~\bibnamefont {Feng}}, \bibinfo {author} {\bibfnamefont {H.}~\bibnamefont
  {Cao}}, \ and\ \bibinfo {author} {\bibfnamefont {S.-W.}\ \bibnamefont
  {Cheong}},\ }\bibfield  {title} {\enquote {\bibinfo {title} {{Bilayer Square
  Lattice ${\mathrm{Tb}}_{2}{\mathrm{SrAl}}_{2}{\mathrm{O}}_{7}$ with
  Structural Z$_8$ Vortices and Magnetic Frustration}},}\ }\href {\doibase
  10.1021/acs.chemmater.1c03771} {\bibfield  {journal} {\bibinfo  {journal}
  {Chemistry of Materials}\ }\textbf {\bibinfo {volume} {34}},\ \bibinfo
  {pages} {1225--1234} (\bibinfo {year} {2022})}\BibitemShut {NoStop}%
\bibitem [{\citenamefont {Ali}\ \emph {et~al.}(2024)\citenamefont {Ali},
  \citenamefont {Xu}, \citenamefont {Bernoudy}, \citenamefont {Nocera},
  \citenamefont {King},\ and\ \citenamefont {Banerjee}}]{Ali2024}%
  \BibitemOpen
  \bibfield  {author} {\bibinfo {author} {\bibfnamefont {A.}~\bibnamefont
  {Ali}}, \bibinfo {author} {\bibfnamefont {H.}~\bibnamefont {Xu}}, \bibinfo
  {author} {\bibfnamefont {W.}~\bibnamefont {Bernoudy}}, \bibinfo {author}
  {\bibfnamefont {A.}~\bibnamefont {Nocera}}, \bibinfo {author} {\bibfnamefont
  {A.~D.}\ \bibnamefont {King}}, \ and\ \bibinfo {author} {\bibfnamefont
  {A.}~\bibnamefont {Banerjee}},\ }\bibfield  {title} {\enquote {\bibinfo
  {title} {{Quantum Quench Dynamics of Geometrically Frustrated Ising
  Models}},}\ }\href {https://arxiv.org/abs/2403.00091} {\bibfield  {journal}
  {\bibinfo  {journal} {arXiv:2023.00091}\ } (\bibinfo {year}
  {2024})}\BibitemShut {NoStop}%
\bibitem [{\citenamefont {Shu}\ \emph {et~al.}(2022)\citenamefont {Shu},
  \citenamefont {Jian},\ and\ \citenamefont {Yin}}]{Shu2022prl}%
  \BibitemOpen
  \bibfield  {author} {\bibinfo {author} {\bibfnamefont {Y.-R.}\ \bibnamefont
  {Shu}}, \bibinfo {author} {\bibfnamefont {S.-K.}\ \bibnamefont {Jian}}, \
  and\ \bibinfo {author} {\bibfnamefont {S.}~\bibnamefont {Yin}},\ }\bibfield
  {title} {\enquote {\bibinfo {title} {Nonequilibrium dynamics of deconfined
  quantum critical point in imaginary time},}\ }\href {\doibase
  10.1103/PhysRevLett.128.020601} {\bibfield  {journal} {\bibinfo  {journal}
  {Phys. Rev. Lett.}\ }\textbf {\bibinfo {volume} {128}},\ \bibinfo {pages}
  {020601} (\bibinfo {year} {2022})}\BibitemShut {NoStop}%
\bibitem [{\citenamefont {Shu}\ and\ \citenamefont {Yin}(2022)}]{Shu2022prb}%
  \BibitemOpen
  \bibfield  {author} {\bibinfo {author} {\bibfnamefont {Y.-R.}\ \bibnamefont
  {Shu}}\ and\ \bibinfo {author} {\bibfnamefont {S.}~\bibnamefont {Yin}},\
  }\bibfield  {title} {\enquote {\bibinfo {title} {Dual dynamic scaling in
  deconfined quantum criticality},}\ }\href {\doibase
  10.1103/PhysRevB.105.104420} {\bibfield  {journal} {\bibinfo  {journal}
  {Phys. Rev. B}\ }\textbf {\bibinfo {volume} {105}},\ \bibinfo {pages}
  {104420} (\bibinfo {year} {2022})}\BibitemShut {NoStop}%
\bibitem [{\citenamefont {Shu}\ \emph {et~al.}(2023)\citenamefont {Shu},
  \citenamefont {Jian}, \citenamefont {Sandvik},\ and\ \citenamefont
  {Yin}}]{Shu2023kz}%
  \BibitemOpen
  \bibfield  {author} {\bibinfo {author} {\bibfnamefont {Y.-R.}\ \bibnamefont
  {Shu}}, \bibinfo {author} {\bibfnamefont {S.-K.}\ \bibnamefont {Jian}},
  \bibinfo {author} {\bibfnamefont {A.~W.}\ \bibnamefont {Sandvik}}, \ and\
  \bibinfo {author} {\bibfnamefont {S.}~\bibnamefont {Yin}},\ }\bibfield
  {title} {\enquote {\bibinfo {title} {{Equilibration of Topological Defects at
  the Deconfined Quantum Critical Point}},}\ }\href
  {https://arxiv.org/abs/2305.04771} {\bibfield  {journal} {\bibinfo  {journal}
  {arXiv:2305.04771}\ } (\bibinfo {year} {2023})}\BibitemShut {NoStop}%
\bibitem [{\citenamefont {Landau}\ and\ \citenamefont
  {Binder}(2009)}]{Landau2009book}%
  \BibitemOpen
  \bibfield  {author} {\bibinfo {author} {\bibfnamefont {D.~P.}\ \bibnamefont
  {Landau}}\ and\ \bibinfo {author} {\bibfnamefont {K.}~\bibnamefont
  {Binder}},\ }\href@noop {} {\emph {\bibinfo {title} {A Guide to Monte Carlo
  Simulations in Statistical Physics}}},\ \bibinfo {edition} {2nd}\ ed.\
  (\bibinfo  {publisher} {Cambridge University Press},\ \bibinfo {year}
  {2009})\BibitemShut {NoStop}%
\bibitem [{\citenamefont {Janssen}\ \emph {et~al.}(2014)\citenamefont
  {Janssen}, \citenamefont {Schaub},\ and\ \citenamefont
  {Schmittmann}}]{Janssen1989}%
  \BibitemOpen
  \bibfield  {author} {\bibinfo {author} {\bibfnamefont {H.~K.}\ \bibnamefont
  {Janssen}}, \bibinfo {author} {\bibfnamefont {B.}~\bibnamefont {Schaub}}, \
  and\ \bibinfo {author} {\bibfnamefont {B.}~\bibnamefont {Schmittmann}},\
  }\bibfield  {title} {\enquote {\bibinfo {title} {New universal short-time
  scaling behaviour of critical relaxation processes},}\ }\href {\doibase
  10.1007/BF01319383} {\bibfield  {journal} {\bibinfo  {journal} {Zeitschrift
  f\"{u}r Physik B Condensed Matter}\ }\textbf {\bibinfo {volume} {73}},\
  \bibinfo {pages} {539} (\bibinfo {year} {2014})}\BibitemShut {NoStop}%
\bibitem [{\citenamefont {Li}\ \emph {et~al.}(1995)\citenamefont {Li},
  \citenamefont {Sch\"ulke},\ and\ \citenamefont {Zheng}}]{Li1995prl}%
  \BibitemOpen
  \bibfield  {author} {\bibinfo {author} {\bibfnamefont {Z.~B.}\ \bibnamefont
  {Li}}, \bibinfo {author} {\bibfnamefont {L.}~\bibnamefont {Sch\"ulke}}, \
  and\ \bibinfo {author} {\bibfnamefont {B.}~\bibnamefont {Zheng}},\ }\bibfield
   {title} {\enquote {\bibinfo {title} {Dynamic monte carlo measurement of
  critical exponents},}\ }\href {\doibase 10.1103/PhysRevLett.74.3396}
  {\bibfield  {journal} {\bibinfo  {journal} {Phys. Rev. Lett.}\ }\textbf
  {\bibinfo {volume} {74}},\ \bibinfo {pages} {3396--3398} (\bibinfo {year}
  {1995})}\BibitemShut {NoStop}%
\end{thebibliography}%

\end{document}